\documentclass[12pt]{article}
\newlength{\pubnumber} 

\catcode`\@=11
\@addtoreset{equation}{section}

\def\section{\@startsection{section}{1}{\z@}{3.5ex plus 1ex minus .2ex}
 {2.3ex plus .2ex}{\large\bf}}
\def\subsection{\@startsection{subsection}{2}{\z@}{2.3ex plus .2ex}
 {2.3ex plus .2ex}{\bf}}

\usepackage{latexsym}

\begin{document}

\begin{titlepage}
\samepage{
\setcounter{page}{0}
\rightline{December 2007}
\vfill
\begin{center}
    {\Large \bf New Regulators for Quantum Field Theories\\ 
     with Compactified Extra Dimensions\\
            {\it --- Part I: ~Fundamentals ---}\\}
\vfill
   {\large Sky Bauman\footnote{ E-mail address:
      bauman@physics.arizona.edu} $\,$and$\,$ Keith
      R. Dienes\footnote{ E-mail address:  dienes@physics.arizona.edu}
      \\}
\vspace{.10in}
 {\it  Department of Physics, University of Arizona, Tucson, AZ  85721
 USA\\}
\end{center}
\vfill
\begin{abstract}
  {\rm  In this paper, we propose two new regulators for quantum
        field theories in spacetimes with compactified extra dimensions.  
        We refer to these regulators as the ``extended hard cutoff'' (EHC)
        and ``extended dimensional regularization'' (EDR).
        Although based on traditional four-dimensional regulators,
        the key new feature of these higher-dimensional regulators is that 
        they are specifically designed to handle mixed spacetimes in which some
        dimensions are infinitely large  and others are compactified. 
        Moreover, unlike most other regulators which have been used in 
        the extra-dimension literature, these regulators are designed to 
        respect the original
        higher-dimensional Lorentz and gauge symmetries that exist prior to
        compactification, and not merely the four-dimensional symmetries which
        remain afterward.  This distinction is particularly relevant for
        calculations of the physics of the excited Kaluza-Klein modes
        themselves, and not merely their radiative effects on zero modes.  By
        respecting the full higher-dimensional symmetries, our regulators
        avoid the introduction of spurious terms which would not have been
        easy to disentangle from the physical effects of compactification. 
         As part of our work, we also derive a number of ancillary results.
        For example, we demonstrate
        that in a gauge-invariant theory, 
        analogues of the Ward-Takahashi identity hold not only for 
        the usual zero-mode (four-dimensional) photons, but for all 
        excited Kaluza-Klein photons as well.}  
\end{abstract}
\vfill
\smallskip}
\end{titlepage}

\setcounter{footnote}{0}

\def\beq{\begin{equation}}
\def\eeq{\end{equation}}
\def\beqn{\begin{eqnarray}}
\def\eeqn{\end{eqnarray}}
\def\half{{\textstyle{1\over 2}}}
\def\quarter{{\textstyle{1\over 4}}}

\def\calO{{\cal O}}
\def\calE{{\cal E}}
\def\calT{{\cal T}}
\def\calM{{\cal M}}
\def\calF{{\cal F}}
\def\calS{{\cal S}}
\def\calY{{\cal Y}}
\def\calV{{\cal V}}
\def\ibar{{\overline{\imath}}}
\def\chibar{{\overline{\chi}}}
\def\ttwo{{\vartheta_2}}
\def\tthree{{\vartheta_3}}
\def\tfour{{\vartheta_4}}
\def\ttwob{{\overline{\vartheta}_2}}
\def\tthreeb{{\overline{\vartheta}_3}}
\def\tfourb{{\overline{\vartheta}_4}}

\def\qbar{{\overline{q}}}
\def\mm{{\tilde m}}
\def\nn{{\tilde n}}
\def\rep#1{{\bf {#1}}}
\def\ie{{\it i.e.}\/}
\def\eg{{\it e.g.}\/}

\newcommand{\newc}{\newcommand}
\newc{\gsim}{\lower.7ex\hbox{$\;\stackrel{\textstyle>}{\sim}\;$}}
\newc{\lsim}{\lower.7ex\hbox{$\;\stackrel{\textstyle<}{\sim}\;$}}

\hyphenation{su-per-sym-met-ric non-su-per-sym-met-ric}
\hyphenation{space-time-super-sym-met-ric}
\hyphenation{mod-u-lar mod-u-lar--in-var-i-ant}


\def\inbar{\,\vrule height1.5ex width.4pt depth0pt}

\def\IC{\relax\hbox{$\inbar\kern-.3em{\rm C}$}}
\def\IQ{\relax\hbox{$\inbar\kern-.3em{\rm Q}$}}
\def\IR{\relax{\rm I\kern-.18em R}}
 \font\cmss=cmss10 \font\cmsss=cmss10 at 7pt
\def\IZ{\relax\ifmmode\mathchoice
 {\hbox{\cmss Z\kern-.4em Z}}{\hbox{\cmss Z\kern-.4em Z}}
 {\lower.9pt\hbox{\cmsss Z\kern-.4em Z}} {\lower1.2pt\hbox{\cmsss
 Z\kern-.4em Z}}\else{\cmss Z\kern-.4em Z}\fi}

\long\def\@caption#1[#2]#3{\par\addcontentsline{\csname
  ext@#1\endcsname}{#1}{\protect\numberline{\csname
  the#1\endcsname}{\ignorespaces #2}}\begingroup \small
  \@parboxrestore \@makecaption{\csname
  fnum@#1\endcsname}{\ignorespaces #3}\par \endgroup}
\catcode`@=12

\input epsf
\section{Introduction
\label{intro}
}
\setcounter{footnote}{0}

Extra dimensions are among the leading candidates for physics beyond the
Standard Model.  However, despite the vast amount of work done in this area,
phenomenological studies of higher-dimensional models still face
limitations.  A fundamental issue is that virtually all realistic
theories in higher dimensions are non-renormalizable.  Because
parameters in a non-renormalizable theory are extremely sensitive to an
ultraviolet (UV) cutoff, and because an infinite number of
counterterms are needed to absorb divergences, our ability to make
meaningful predictions at different energy scales appears to be
compromised.  Additionally, regulators of UV divergences can introduce
unphysical artifacts.  For example, a hard cutoff in QED artificially
generates a large photon mass term.  The problem of artifacts should be
especially severe in higher-dimensional theories since
the non-renormalizability will magnify any such radiative effect.

Unfortunately, such artifacts will be introduced by many of
the regulators which are typically used to perform calculations
in spacetimes with compactified extra dimensions.
This happens because these regulators artificially treat
momentum components along compactified extra dimensions as if they
were separate from the other components.  To be more explicit, let
us consider a typical one-loop diagram in a theory with
a single universal compactified extra dimension.
The amplitude corresponding to such a diagram 
can be expressed as a mode-number sum over a four-momentum
integral, \ie, 
\beq
\mathcal{M} ~=~ \sum_r \, \int \, \frac{d^4 k}{(2\pi)^4} ~ I(k,r)~,
\label{sum_int}
\eeq
where $\mathcal{M}$ represents the one-loop amplitude, $k$ is the
four-momentum of a Kaluza-Klein (KK) state running in the loop, and
$r$ is its KK mode number.  The function $I$ depends on $k$ and $r$, 
as well as the the couplings in the theory  
and momenta and mode numbers of any external particles. 
Of course, both the four-momentum integral and the KK sum
contribute to possible divergences, and both potential
sources of infinities must be regularized. 

The typical approach is to apply a standard four-dimensional regulator 
(such as a hard cutoff or dimensional regularization) to the integral, 
and to truncate the sum at large but finite limits.
Thus, the sum and the integral are regulated independently. 
Unfortunately, independent regularizations artificially 
violate the higher-dimensional Lorentz
invariance that originally existed in the theory, 
thereby leading to unphysical artifacts in $\mathcal{M}$.  
This is because the variables $k$ and $r/R$ from Eq.~(\ref{sum_int}) 
are actually part of a single five-momentum running in the loop. 
Our regulator should therefore reflect this
higher-dimensional Lorentz symmetry,
just as a hard cutoff in four dimensions (4D) is always imposed on the 
total Euclidean four-momentum running in a loop, and not a particular subset
of momentum components. This is why separate
regularizations of four-momentum integrals and KK sums violate higher-dimensional 
Lorentz invariance. 
Without respecting the full five-dimensional Lorentz symmetry,
any such regulator has the potential to introduce unphysical artifact terms 
into the results of any calculation.

Of course, it might seem that
we can always subtract unphysical artifacts at the end of a
calculation.  However, this is not generally possible because the compactification
itself, which breaks the higher-dimensional Lorentz invariance globally,
can also induce local violations of the higher-dimensional
Lorentz invariance in an effective field theory (EFT) at finite
energy.
We would therefore not know which terms to subtract, since
it would be extremely difficult to distinguish these unphysical artifacts from
the expected {\it bona-fide}\/ violations
of five-dimensional Lorentz invariance stemming which arise due to the compactification.

Given this situation,
our goal in this paper is to develop a set of regulators
which are based firmly on two
fundamental higher-dimensional symmetries:
\begin{itemize}
\item higher-dimensional Lorentz invariance;  and
\item higher-dimensional gauge invariance, when appropriate.
\end{itemize}
Regulators which are based on these symmetries should therefore
be broadly applicable and free of unphysical artifacts.
Moreover, we shall also require that our regulators be
 {\it theory-independent}\/.
In other words, we shall require that our regulators be 
insensitive to the specific particle content and interactions characterizing the
field theory in question. 

In this paper, we shall develop two distinct regulator schemes 
which meet these criteria.
Indeed, in each case, these regularization methods can be 
viewed as higher-dimensional generalizations of 
well-known four-dimensional regulators.
However, as discussed above,
their distinguishing property is that they control
four-momentum integrals and KK sums {\it collectively}\/, as appropriate for
higher-dimensional calculations. 
Under this scheme,
the constraints on the integral and the sum in Eq.~(\ref{sum_int}) 
become coupled.

Our first regulator
will be a generalization of a four-dimensional hard-cutoff scheme
to the case of theories with KK modes. 
We shall refer to this regulator as an ``extended hard cutoff'' (EHC) regulator. 
To do this, we shall consider the case of a
single extra dimension compactified on a circle.  Instead of separately
regulating four-momentum integrals and KK sums, we shall implement a cutoff 
on the {\it total five-momenta}\/ of virtual KK states running through internal
loops.  This procedure is Lorentz-invariant, and therefore does not
introduce unphysical artifacts.

By contrast,
our second regulator will be a generalization of dimensional regularization,
to be referred to as ``extended dimensional regularization'' (EDR).
Specifically, we shall use standard dimensional-regulatorization techniques 
to control four-momentum integrals.
However, we shall also demand that KK sums be
truncated at limits which depend on the dimensional regularization
parameter $\epsilon$. 
The critical point, then, is to determine an appropriate balancing relation between
this KK cutoff and the dimensional-regularization parameter $\epsilon$ 
which preserves not only higher-dimensional
Lorentz invariance, but also higher-dimensional gauge invariance.
To do this, we shall consider the case of  
five-dimensional QED compactified on a circle
and show explicitly that 
preserving both higher-dimensional Lorentz invariance and gauge invariance
in this theory
leads to a unique relation 
between $\epsilon$ and the KK cutoff.
Our criterion of theory-independence will then guarantee that this 
relation between the KK cutoff and $\epsilon$ should hold for {\it all}\/ 
higher-dimensional field theories, regardless of whether or 
not they contain gauge symmetries.

At this stage, one might be tempted to offer two possible objections
to the approach we shall be following in this paper.
First, since 
the compactification itself distinguishes extra dimensions from the ones 
we currently observe, one could argue that
there is no need to respect higher-dimensional
Lorentz invariance.
Indeed, one might even argue that the very process of compactification forces
us to employ regulators that do {\it not}\/ respect higher-dimensional Lorentz
invariance:  since the momentum components along compactified dimensions
are discrete variables and components along large dimensions are
continuous, it might seem that no regularization scheme can truly 
put these variables on equal footing. 
However, it is important to realize that compactification 
is an effect at finite distance and therefore finite energies. 
In the UV limit, by contrast, this discreteness fades away and 
higher-dimensional Lorentz invariance is restored. 
Since regulators are designed to control UV divergences, it is therefore
essential that they respect whatever symmetries exist at short distances.

Second, one might object that higher-dimensional
theories are non-renormalizable.  Therefore, it would seem that
we should obtain meaningless results regardless of which regulator we use, 
in which case there 
is no point in trying to extract exact predictions from such
theories.  However, despite the non-renormalizability, 
it is possible to derive precise, {\it finite}\/ relationships between 
the {\it renormalized}\/ parameters in our effective field theories 
that characterize KK states. 
Indeed, as we shall explicitly demonstrate in Ref.~\cite{paper1b},
the use of proper regulators will allow us to relate
the parameters describing excited  KK modes to the corresponding parameters
describing zero modes, after
each have received radiative corrections. 
If the zero-mode parameters are taken to be experimental inputs, then 
the entire KK spectrum is determined.  We emphasize that this only 
works when regulators are designed to respect higher-dimensional symmetries.

Although our extended hard-cutoff and extended dimensional-regularization procedures
ultimately achieve the same goal, there are two significant 
conceptual differences between them.
First, our extended hard cutoff is designed to treat all components of loop momenta
in the same way, and hence this cutoff never breaks higher-dimensional Lorentz
invariance.  By contrast, our extended dimensional-regularization procedure
controls divergences from four-momentum integrals and KK sums through
very different means.  Higher-dimensional symmetries thus do not appear
to be preserved from the outset, but survive in the end only because of a special
relation between their regularization parameters.

There also is a second important difference.
Because a hard cutoff explicitly violates gauge invariance,
our extended hard-cutoff regulator will not be suitable for higher-dimensional theories 
in which gauge symmetries are present.
By contrast, our extended dimensional-regularization procedure is designed
to respect higher-dimensional gauge invariance as well
as higher-dimensional Lorentz invariance.  As such, this is
the regulator of choice when dealing with gauge-invariant theories.
In this connection, we remark that while the  process of compactification
explicitly violates Lorentz invariance globally (and this can
translate into local Lorentz violations below the UV limit),
the process of compactification in and of itself does {\it not}\/
violate any higher-dimensional gauge symmetry which exists in the UV limit. 
Specifically, as we shall demonstrate for the case of 
five-dimensional QED compactified on a circle,
a full five-dimensional gauge invariance survives after compactification,
even at low energy scales.
Our extended dimensional-regularization procedure will reflect this explicitly
through the preservation of Ward identities and Ward-Takahashi identities;
indeed, such identities will continue to hold not only for the (zero-mode) photon,
but for all of the excited (KK) photons as well.

This paper is organized as follows. In Sect.~\ref{hard}, we introduce
our extended hard cutoff (EHC) regulator, and explain how it regularizes divergences in a
Lorentz-invariant fashion. In Sect.~\ref{dimregsect}, we then introduce
our extended dimensional regulator (EDR) in the context of higher-dimensional gauge
theory.  In Sect.~\ref{need}, we turn to a discussion of other
regulators which have been utilized in the literature, 
and compare our regulators with those.  We also demonstrate, 
through explicit examples,
the kinds of difficulties that
can arise when one uses a regulator which does not respect
higher-dimensional Lorentz invariance.  We also discuss
the relations between our EHC and EDR regulators and 
several other Lorentz-invariant methods which have already been 
developed in the literature.
Finally, Sect.~\ref{conclude} contains our conclusions and ideas
for possible extensions.

This paper is the first in a two-part series.
In this paper, we shall focus on the development of two new regulators,
as sketched above.  By contrast, in a subsequent companion paper~\cite{paper1b},
we shall discuss how these new regulators may be employed
in order to derive effective field theories at different energy scales.
We shall also discuss how these regulator techniques can be used
to extract finite results for physical observables that
relate the physics of excited KK modes to the physics of KK zero modes.
In this context, it should be noted that one of our 
primary motivations for developing these new EHC and EDR regulators
has been to enable us to study the way in which 
the Kaluza-Klein mass and coupling parameters in any higher-dimensional
effective field theory evolve as a function of energy scale.
For example, we might wish to study how the well-known tree-level
relations amongst the tower of KK masses and amongst their couplings
are ``deformed'' when radiative effects are included.
This will be the subject of a third paper~\cite{paper2}.  
However, each of these subsequent papers will rely on the
regulators and calculational techniques that we shall be developing here.

\section{The Extended Hard Cutoff (EHC) Regulator\label{hard} }
\setcounter{footnote}{0}

In this section, we introduce our higher-dimensional extended hard cutoff (EHC) 
regulator.
For simplicity, we consider the case of a single extra dimension compactified on a 
circle;  generalizations to other compactifications will be
straightforward.  As discussed in the Introduction, our cutoff will be purely 
five-dimensional in nature, and will respect five-dimensional Lorentz invariance 
from the outset.
Of course, if our higher-dimensional theory is also gauge invariant, then
a hard cutoff will not be applicable;  in such cases, the EDR regulator in 
Sect.~3 should be used.

\begin{figure}[ht]
\centerline{
   \epsfxsize 3.0 truein \epsfbox {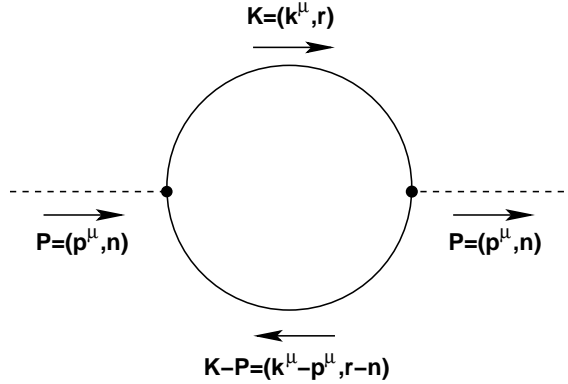} }
\caption{A generic one-loop diagram:  
   an external Kaluza-Klein particle (dotted line)  with four-momentum
   $p^\mu$ and Kaluza-Klein index $n$ interacts with a tower of
   Kaluza-Klein particles (solid lines) of bare mass $M$.  }
\label{fig1}
\end{figure}

To illustrate our procedure, let us consider a generic one-loop
diagram of the form shown in Fig.~\ref{fig1} in which an external
particle with four-momentum $p^\mu$ and mode number $n$ interacts with
a tower of KK particles  of bare mass $M$.  Enforcing 5D momentum
conservation at the vertices (as appropriate for compactification on a
circle) 
and assuming that the solid lines correspond to scalar fields
leads to a one-loop integral of the form
\beq
  L_n (p) ~=~ \sum_r \int \frac{d^4 k}{(2\pi)^4} ~ \frac{1}{k^2 - r^2
          /R^2 - M^2} ~\frac{1}{(k - p)^2 - (r - n)^2 /R^2 - M^2}~
\label{gral}
\eeq
where $k$ is the four-momentum of a particle in our loop and $r$ is
its mode number.  Although we are considering a particular form for a
loop integral,  we will keep our discussion as general as possible.

Following standard techniques, we may immediately rewrite this loop
integral  as
\beq
   L_n (p) ~=~ i\int_{0}^{1}dx\, \sum_r \, \int \frac{d^4
               \ell_E}{(2\pi)^4} \left[\frac{1}{\ell_{E}^2 +
               {\ell^4}^2 + \mathcal{M}^2 (x)}\right]^2 ~,
\label{intwick}
\eeq
where $x$ is a Feynman parameter, where $\ell$ represents the shifted
momentum
\beq
            \ell ~\equiv~  k - xp~,~~~~~ \ell^4 ~\equiv~ (r - xn)/R~,
\label{ldef}
\eeq
where $\ell_E$ is the Euclidean (Wick-rotated) momentum
\beq
              \ell^{0}_{E} ~\equiv~ -i\ell^0 ~,~~~~~ \vec \ell_{E}
              ~\equiv~ \vec \ell~ ,
\eeq
and where the effective mass in Eq.~(\ref{intwick}) is given by
\beq
         \mathcal{M}^2 (x) ~\equiv~ M^2 ~+~ x(x - 1)\left\lbrack p^2 -
         {n^2\over R^2} \right\rbrack~.
\label{delt}
\eeq
Note that throughout this paper, vector and tensor components corresponding 
to the fifth dimension will be denoted with a superscript `4'.  We have
chosen this somewhat unorthodox convention in order to emphasize
the preservation of five-dimensional Lorentz invariance, so that
our five-dimensional Lorentz indices are given as $M=0,1,2,3,4$.

We now introduce our hard momentum cutoff $\Lambda$.  We shall apply
this directly to the Euclidean {\it five}\/-momentum running in the loop, as
appropriate for an intrinsically five-dimensional calculation, so that
\beq
            \ell_{E}^2 + {\ell^4}^2 ~\leq~ \Lambda^2 ~.
\label{cutoff}
\eeq
Of course, this constraint equation
correlates the cutoff for the integration  over the
four-momentum $\ell_E$ with the cutoff for the summation over the KK
index $r$.  In particular, the constraint in Eq.~(\ref{cutoff}) can be
implemented by restricting the KK summation to integers in the range
\beq
           -\Lambda R + xn ~\leq~ r ~\leq~ \Lambda R + xn~
\label{rcut}
\eeq
and then restricting our $\ell_E$-integration to the  corresponding
range
\beqn
            \ell_{E}^2 ~& \leq&~  \Lambda^2 - {\ell^4}^2 \nonumber\\ &
                        \leq&~  \Lambda^2 - (r-xn)^2/R^2~.
\label{kcut}
\eeqn

Clearly, Eqs.~(\ref{rcut}) and~(\ref{kcut}) are nested constraints on the
components of the momentum of the particle running in the loop.
However, this ``nesting'' is unavoidable
if our regulator is to preserve five-dimensional Lorentz invariance and
avoid distinguishing a special direction in spacetime.
Since $\ell_E$ is continuous
and $\ell^4$ is discrete, one might argue at first glance
that these variables are
fundamentally different, and that Eq.~(\ref{cutoff}) does not truly
respect a five-dimensional Lorentz symmetry. However, as discussed in the Introduction,
the discreteness is an effect at {\it finite}\/ energy scales, originating from 
the compactification.  This discreteness is
not apparent in the UV limit, where the gaps between KK masses are
effectively negligible. Therefore, Eq.~(\ref{cutoff}) will indeed allow us to
regularize five-dimensional UV divergences in a Lorentz-invariant fashion.

Eqs.~(\ref{cutoff}) through (\ref{kcut}) define our extended hard-cutoff (EHC) 
regularization
procedure.  Indeed, unlike the case with dimensional regularization
to be discussed in Sect.~3, the maintenance of five-dimensional
Lorentz invariance in this case has not been particularly difficult or profound.
However, this is not enough, since we also need to know how to
perform calculations which implement these constraints.  Eq.~(\ref{rcut})
is particularly unpleasant, since it puts the Feynman parameter and
the mode number of the external particle in the summation limits. One
might hope that we can neglect these terms when $\Lambda$ is
large. However this is ultimately not possible due to the hypersensitivity to the exact
value of a cutoff in a non-renormalizable theory. The rest of this
section is therefore devoted to the calculational issue of converting 
such expressions for loop diagrams into useful forms.

For the special case of $n = 0$, our loop diagram can be written as
\beq
           L_0 (p) ~=~ i \int_{0}^{1}dx \, \sum_{r = -\Lambda
           R}^{\Lambda R}\, f_0 (p,r,x)
\label{f0}
\eeq
where $f_0$ is the integral over $\ell_E$ from Eq.~(\ref{intwick}),
subject to the constraint in Eq.~(\ref{kcut}).  In general, $f_0$ is a
function of $p$, $r$, and $x$, but we will not need to evaluate $f_0$
for this discussion.  Note that in writing Eq.~(\ref{f0}), we have
treated $\Lambda R$ as an integer.  In the limit of a large cutoff,
this assumption will have no effect on our results.

For $n\not=0$, however, the cutoff on the $r$-summation depends on the
Feynman parameter $x$.  Fortunately, we can eliminate this dependence 
through a series of variable substitutions.  Let us first assume that $n>0$. In
this case, our summation is  over all integers $r$ in the range
$-\Lambda R+xn \leq r \leq \Lambda R+xn$.  In the following, we shall
adopt a notation whereby $\sum_{r=a}^b$ denotes a summation over
integer values of $r$ within the range $a\leq r\leq b$ even if $a$ and
$b$ are not themselves integers.  We can then write
\beq
L_n (p) = i\int_{0}^{1}dx \sum_{r = -\Lambda R + xn}^{\Lambda R +
xn}f_n (p,r,x)~,
\label{L_n}
\eeq
where $f_n$ is the analog of $f_0$ for non-zero $n$.  For $n > 0$, we
may express this summation as
\beqn
           \int_0^1 dx\, \sum_{r=-\Lambda R+xn}^{\Lambda R+xn}  ~&=&~
         {1\over n} \int_0^n du \, \sum_{r=-\Lambda R+u}^{\Lambda R+u}
         ~~~~~~~~~~~ {\rm where}~~ u\equiv xn  \nonumber\\ ~&=&~
         {1\over n} \sum_{j=0}^{n-1} \, \int_j^{j+1} du \,  \sum_{r=
         -\Lambda R+u}^{\Lambda R+u} \nonumber\\ ~&=&~ {1\over n}
         \sum_{j=0}^{n-1} \, \int_0^{1} d\hat u \,  \sum_{\hat r =
         -\Lambda R + \hat u}^{\Lambda R+\hat u}  ~~~~~~ {\rm where}~~
         \cases{  \hat u \equiv u-j & ~\cr \hat r \equiv r-j & ~\cr}
         \nonumber\\ ~&=&~ {1\over n} \sum_{j=0}^{n-1} \, \int_0^{1}
         d\hat u \,  \sum_{\hat r = -\Lambda R + 1}^{\Lambda R} ~.
\label{manysteps}
\eeqn
In passing to the last line, we have continued to treat $\Lambda R$ as an
integer.  We have also used the fact that the exact $\hat u=\lbrace 0,1\rbrace$
endpoints of the $\hat u$-integration region are sets of measure zero.

For general $n\not=0$ of either sign, we can make an analogous set  of
substitutions, resulting in the general identity:
\beq
           \int_0^1 dx\, \sum_{r=-\Lambda R+xn}^{\Lambda R+xn}  ~=~
         {1\over |n|} \sum_{j=0}^{|n|-1} \, \int_0^{1} d\hat u \,
         \sum_{\hat r = -\Lambda R + 1}^{\Lambda R}~,
\label{kgenid}
\eeq
where
\beq
           \hat u ~\equiv~ x|n|-j
\label{u_hat}
\eeq
and
\beq
             \hat r ~\equiv~ {\rm sign}(n) r - j~.
\label{r_hat}
\eeq
Together, Eqs.~(\ref{u_hat}) and~(\ref{r_hat}) imply that 
$R\ell^4 = {\rm sign}(n)[\hat r -\hat u]$. 

Note that the mode number $n$ has
disappeared from the magnitude of $\ell^4$. 
This is precisely as we expect, since $\ell^4$ is merely a summation 
variable and should not depend on the magnitude of $n$. 
Likewise,
the dependence on ${\rm sign}(n)$ arises by convention and can be absorbed
into coefficients of $\ell^4$. 
Ultimately, this removal of $n$ from $\ell^4$ 
was possible only because of the limits we chose for $r$ at the beginning
of our calculation. 

However, it is important to realize that $n$ has not vanished from our calculation. 
Because $\hat{u}$ is now viewed as an independent Feynman-like variable
in Eq.~(\ref{kgenid}), 
$x$ must now be expressed in terms of $\hat{u}$, and this reintroduces
a dependence on $n$ into any expressions which previously depended on $x$.
For example, the quantity $x$
appears within $\mathcal{M}^2$, as defined in Eq.~(\ref{delt}). 
However, the important point is that this dependence on $n$ is now wholly within
the four-dimensional integrand, and no longer appears within the KK summation limits.

Given these variable substitutions,
our loop-diagram expression for non-zero $n$ now be rewritten as
\beq
       L_n (p) ~=~ i\int_{0}^{1}d\hat u \, \frac{1}{|n|}\, \sum_{j =
            0}^{|n| - 1} \, \sum_{\hat r = -\Lambda R + 1}^{\Lambda R}
            \, f_n (p,\hat r,\hat u,j)~.
\label{fn}
\eeq
We shall henceforth drop the hats from $\hat r$ and $\hat u$.
Note that the functions $f_0 (r,u)$ and $f_n (r,u,j)$  each depend on
the cutoff $\Lambda$ because  they are integrals whose limits contain
$\Lambda$. For example, if our original diagram is of the form
(\ref{intwick}), then these functions $f_0$ and $f_n$ are given by
\beqn
   f_0(r,u) &=& \int {d^4\ell_E\over (2\pi)^4} \, \left\lbrack {1\over
                  \ell_E^2 + r^2/R^2+ M^2 + u(u-1)p^2 }\right\rbrack^2
                  \nonumber\\ f_n(r,u,j) &=& \int {d^4\ell_E\over
                  (2\pi)^4} \, \left\lbrack {1\over \ell_E^2 +
                  (r-u)^2/R^2+ M^2 + (u+j)(u+j-|n|) \left({p^2\over
                  n^2} - {1\over R^2}\right)}\right\rbrack^2\nonumber\\
\label{fexamples}
\eeqn
where these integrals are subject to the cutoffs
\beqn
   f_0:~~~~~~  \ell_E^2 & \leq&  \Lambda^2 - r^2/R^2\nonumber\\
   f_n:~~~~~~  \ell_E^2 & \leq& \Lambda^2 - (r-u)^2/R^2
\eeqn
respectively.  Note that $f_n$ is the same as $f_0$, but with the
simultaneous algebraic substitutions $r\to \rho\equiv r-u$, $u\to
y\equiv (u+j)/|n|$, and $p^2 \to p^2 - n^2 /R^2$.

Eqs.~(\ref{f0}) and~(\ref{fn}) are the main results of this
section.  Once loop diagrams are in these forms, they can be evaluated
directly using standard four-dimensional techniques. 
Similarly, although we restricted ourselves to the case of a single external
particle, generalizations to more complicated
situations are straightforward.

Finally, before concluding our discussion of our EHC regulator,
we remark that the identity we have outlined in Eq.~(\ref{kgenid})
relies rather fundamentally on the assumption that the one-loop diagrams
we are regulating can be evaluated through the introduction of only
a single Feynman parameter $x$ (or $u$).
However, this procedure readily generalizes to diagrams that would utilize
arbitrary numbers of Feynman parameters.

\begin{figure}[ht]
\centerline{
   \epsfxsize 3.0 truein \epsfbox {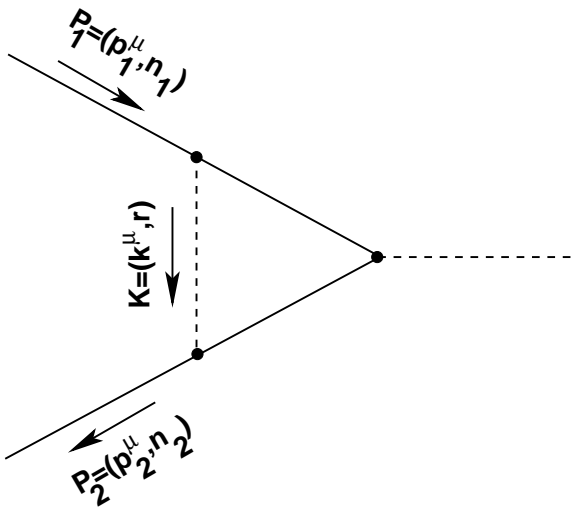} }
\caption{A generic one-loop diagram with three external particles
and three internal propagators.  Such a one-loop diagram will 
require two Feynman parameters.}
\label{fig2}
\end{figure}

As a concrete example, let us consider a diagram such as the
one-loop vertex correction in Fig.~\ref{fig2}
which would require two Feynman parameters. 
In general, such a diagram will take the algebraic form
\beq
    L_{n_1,n_2}(p_1,p_2) ~=~ 
      i\int_{0}^{1}dx_1 \int_{0}^{1}dx_2 \, \sum_r \, f_{n_1,n_2}(p_1,p_2,r,x_1,x_2)~
\label{two_feyn}
\eeq
where $x_1$ and $x_2$ are our two Feynman parameters and
$f$ is our four-momentum integral. 
However, unlike the case of a single Feynman parameter,
our shifted momentum within $f$
will now be given by
\beq
    \ell ~\equiv~ k - x_1 p_1 - x_2 p_2~,~~~~~~
    \ell^4 ~=~ (r - x_1 n_1 - x_2 n_2)/R~.
\label{2ldef}
\eeq
Despite this change in the definition of $\ell$,
our EHC regularization condition 
continues to take the same form as in Eq.~(\ref{cutoff}).
The cutoffs on our KK summation therefore now take the form
\beq
    -\Lambda R +x_1 n_1 + x_2 n_2 ~\leq~ r ~\leq ~\Lambda R + x_1 n_1 + x_2 n_2~
\label{rstraint}
\eeq
while our corresponding four-momentum integral is subject to the cutoff
\beq
         \ell_E^2 ~\leq~ \Lambda^2 - (r-x_1n_1-x_2n_2)^2/R^2~. 
\label{kintcutoff}
\eeq

As before, the primary difficulty here is the presence of the 
Feynman parameters $x_1$ and $x_2$ in the upper and lower limits of the KK 
summation in Eq.~(\ref{2ldef}).
However, these can be eliminated in a manner completely analogous to the method
outlined in Eq.~(\ref{kgenid}).
First, we observe that when $n_1=n_2=0$,
the Feynman parameters are eliminated trivially,
and  Eq.~(\ref{rstraint}) reduces to $-\Lambda R \leq r \leq \Lambda R$. 
Moreover, when one $n_i=0$ but the other is non-zero,
only one Feynman parameter appears in Eq.~(\ref{rstraint}). 
The variable transforms introduced in Eq.~(\ref{kgenid}) 
may therefore be employed to disentangle the remaining Feynman parameter 
from the summation limits. 
As a result, the only new non-trivial case is the one 
in which both $n_1$ and $n_2$ are non-zero. 

Let us first consider the case in which 
both $n_1$ and $n_2$ are positive.  Repeating the steps 
in Eq.~(\ref{manysteps}),
we can then write
\beqn
&&  \int_{0}^{1}dx_1 \int_{0}^{1}dx_2 
       \sum_{r = -\Lambda R + x_1 n_1 + x_2 n_2}^{\Lambda R + x_1 n_1 + x_2 n_2}
                      ~=~\nonumber\\  
&& ~~~~ =~\frac{1}{n_1 n_2}\int_{0}^{n_1}du_1 \int_{0}^{n_2}du_2 
       \sum_{r = -\Lambda R + u_1 + u_2}^{\Lambda R + u_1 + u_2} 
             ~~~~~~~~~~~~~~~~ {\rm where}~~ u_i\equiv x_i n_i \nonumber \\
&& ~~~~ =~\frac{1}{n_1 n_2} \sum_{j_1 = 0}^{n_1 - 1} \,\sum_{j_2 = 0}^{n_2 - 1}
            \int_{j_1}^{j_1 + 1}du_1 \int_{j_2}^{j_2 + 1}du_2 
       \sum_{r = -\Lambda R + u_1 + u_2}^{\Lambda R + u_1 + u_2}\nonumber\\ 
&& ~~~~ =~\frac{1}{n_1 n_2} \sum_{j_1 = 0}^{n_1 - 1} \, \sum_{j_2 = 0}^{n_2 - 1}
            \int_{0}^{1}d\hat{u}_1 \int_{0}^{1}d\hat{u}_2 
 \sum_{\hat{r} = -\Lambda R + \hat{u}_1 + \hat{u}_2}^{\Lambda R + \hat{u}_1 + \hat{u}_2}
             ~~~~~~~ {\rm where}~~ \cases{
             \hat u_i\equiv u_i-j_i & ~\cr
             \hat r\equiv r-j_1-j_2~.&~\cr}\nonumber\\
&& ~~~~ =~\frac{1}{n_1 n_2} \sum_{j_1 = 0}^{n_1 - 1}\, \sum_{j_2 = 0}^{n_2 - 1}
            \int_{0}^{1}d\hat{u}_1 
      \left\lbrack
     \int_{0}^{1-\hat u_1}d\hat{u}_2 \sum_{\hat{r} = -\Lambda R + 1}^{\Lambda R}
     + \int_{1-\hat u_1}^{1}d\hat{u}_2 \sum_{\hat{r} = -\Lambda R + 2}^{\Lambda R+1}
         \right\rbrack~.
\label{two_feyn2}
\eeqn
In passing to the final line, we have continued to treat $\Lambda R$ as an integer
and recognized that while the
combination $\hat{u}_1 + \hat{u}_2$ ranges from $0$ to $2$, the 
summation index $\hat{r}$ ranges over the following values:
 \beq
\cases{
   -\Lambda R + 1 \leq \hat{r} \leq \Lambda R & ~~for $0 < \hat{u}_1 + \hat{u}_2 < 1$ \cr
   -\Lambda R + 2 \leq \hat{r} \leq \Lambda R + 1 & ~~for $1 < \hat{u}_1 + \hat{u}_2 < 2$~.}
\label{r_range}
\eeq
Dropping the hats, it follows that under the EHC regulator,
the diagram $L_{n_1,n_2}(p_1,p_2)$ in Eq.~(\ref{two_feyn}) 
with $n_1,n_2>0$ can be rewritten as
\beqn
    && L_{n_1,n_2}(p_1,p_2)  ~=~
      i\int_{0}^{1}dx_1 \int_{0}^{1}dx_2 \, \sum_r \, 
          f_{n_1,n_2}(p_1,p_2,r,x_1,x_2)~\nonumber\\
  && ~~~~=~ \frac{i}{n_1 n_2} \sum_{j_1 = 0}^{n_1 - 1}\, \sum_{j_2 = 0}^{n_2 - 1}
            \int_{0}^{1}d{u}_1 
      \left\lbrack
     \int_{0}^{1- u_1}d{u}_2 \sum_{{r} = -\Lambda R + 1}^{\Lambda R}
     + \int_{1- u_1}^{1}d{u}_2 \sum_{{r} = -\Lambda R + 2}^{\Lambda R+1}
         \right\rbrack \, f_{n_i} (p_i, r, u_i, j_i)\nonumber\\   
  && ~~~~=~ \frac{i}{n_1 n_2} \sum_{j_1 = 0}^{n_1 - 1}\, \sum_{j_2 = 0}^{n_2 - 1}
            \int_{0}^{1}d{u}_1 
            \int_{0}^{1}d{u}_2 
     \sum_{{r} = -\Lambda R }^{\Lambda R}
          f_{n_i} (p_i, r, u_i, j_i) ~+~ E 
\label{Lform}
\eeqn
where $\ell^4=(r-u_1-u_2)/R$,
where the one-loop integrals $f_{n_1,n_2}$ are regulated
according to Eq.~(\ref{kintcutoff}),
and where $E$ denotes an ``endpoint contribution'' which depends on the
particular values of $f$ at or near the cutoff endpoints of the KK summation,
as given below.

The above results are given for the case in which $n_1$ and $n_2$ are both positive.
However, we can handle the general case in which both $n_1$ and $n_2$ are non-zero
as follows.
Let us define $s_i\equiv {\rm sign}(n_i)$, and likewise let us define
$\hat u_i \equiv x_i |n_i| - j_i$ and 
$\hat r\equiv r - s_1 j_1 - s_2 j_2$.
Note that in terms of these variables, we have
$R\ell^4 = \hat r - s_1 \hat u_1 - s_2 \hat u_2$.
Dropping the hats, we then find the general identity
\beqn
     L_{n_1,n_2}(p_1,p_2)  ~=~ 
   \frac{i}{|n_1 n_2|} \sum_{j_1 = 0}^{|n_1| - 1}\, \sum_{j_2 = 0}^{|n_2| - 1}
            \int_{0}^{1}d{u}_1 
            \int_{0}^{1}d{u}_2 
     \sum_{{r} = -\Lambda R }^{\Lambda R}
          f_{n_i} (p_i, r, u_i, j_i) ~+~ E_{s_1,s_2}\nonumber\\ 
\label{Lformgen}
\eeqn
where the endpoint contributions $E_{\pm,\pm}$ are given as 
\beq
  E_{s_1,s_2} ~=~ -\frac{i}{|n_1 n_2|} \sum_{j_1 = 0}^{|n_1| - 1}\, \sum_{j_2 = 0}^{|n_2| - 1}
            \int_{0}^{1}d{u}_1   \, \hat E_{s_1,s_2}
\eeq
with
\beqn
     \hat E_{++} &\equiv& 
            \int_{0}^{1}d{u}_2  \, f(-\Lambda R)  +             
            \int_{1-u_1}^{1}d{u}_2  \,\left[   f(-\Lambda R+1) - f(\Lambda R+1) \right]~\nonumber\\
     \hat E_{--} &\equiv& 
            \int_{0}^{1}d{u}_2  \, f(\Lambda R)  +             
            \int_{1-u_1}^{1}d{u}_2  \,\left[   f(\Lambda R-1) - f(-\Lambda R-1) \right]~\nonumber\\
     \hat E_{+-} &\equiv& 
            \int_0^{u_1} du_2\,  f(-\Lambda R) 
          + \int_{u_1}^1 du_2\,  f( \Lambda R) \nonumber\\ 
     \hat E_{-+} &\equiv& 
            \int_0^{u_1} du_2\,  f( \Lambda R) 
          + \int_{u_1}^1 du_2\,  f(-\Lambda R)~. 
\label{edefs}
\eeqn
In writing Eq.~(\ref{edefs}, we have suppressed all indices and variables for the $f$-functions except 
their dependence on the KK mode number $r$. 
These results have obvious generalizations to one-loop diagrams with 
additional Feynman parameters.

We see, then, that our EHC regulator is quite general,
and that the methods outlined here enable us to 
eliminate the resulting Feynman parameters from the upper and lower
limits on our KK summations.

\section{Extended Dimensional Regularization (EDR)
\label{dimregsect}}

In this section, we turn to our 5D extended dimensional-regularization (EDR)
procedure.
Unlike the case of the hard cutoff in Sect.~2, our  
extended dimensional-regularization procedure is designed to respect
not only five-dimensional Lorentz invariance, but also five-dimensional
gauge invariance.
As discussed in the Introduction, this 
will happen as the result of a careful balancing between the dimensional
regularization parameter $\epsilon$ which regulates the four-dimensional
momentum integral and the KK cutoff $\Lambda$ 
which regulates the KK sum. 

This section is organized as follows.
We start with a preliminary exposition of
our procedure in Sect.~\ref{prelim}. 
Then, in Sect.~\ref{ward},
we discuss the method by which gauge invariance is maintained
by demonstrating that the Ward(-Takahashi) identities must hold not only for
the zero-mode photon, but also for all KK excitations of the photon. 
In Sect.~\ref{sumstraint}, we then use this in order to generate
a relation between the cutoff  parameters
used for the momentum integrals and
the KK mode-number sums.
Finally, in Sect.~\ref{sumstraintexact}, we deal with a number
of loose ends.   For example, we show
that this relation 
implies that five-dimensional Lorentz invariance will be preserved as well.

\subsection{Preliminary steps
\label{prelim}}

We begin by 
considering a generic one-loop amplitude in five dimensions, with one dimension
compactified on a circle.  As with the diagram in Fig.~\ref{fig1}, we will
assume that we have a certain fixed number of external particles with 
four-momenta $p_i^\mu$
and KK mode numbers $n_i$ which enter the diagram as initial states
or exit as final states.  
We shall also assume that only one Feynman parameter is needed;
the generalization to multiple Feynman parameters is straightforward.
Such an amplitude then generally takes the form
\beq
  L_n^{MN...} (p_1,p_2,...) ~=~ i\int_{0}^{1}dx\, \sum_r \,  
            \int \frac{d^4\ell_E}{(2\pi)^4} 
                  ~ \Omega_n^{MN...}(\ell_E,r,x) 
\label{kpredimwick}
\eeq
where $\Omega_n^{MN...}(\ell_E,r,x)$ is an appropriate unspecified integrand
and where the overall $n$ subscript denotes the collection of external KK indices.
Here $M,N,...$ are five-dimensional Lorentz indices appropriate for the
diagram in question;  thus, unlike the situation in Sect.~2, we are now explicitly
indicating that these amplitudes need not be Lorentz scalars.
We shall also assume that our theory contains a five-dimensional gauge
invariance prior to compactification.

We now seek to develop a regularization procedure
for such amplitudes which is based on the traditional 't~Hooft-Veltman dimensional
regularization procedure~\cite{dimreg} for the four-momentum integral.  
However, we need to regulate not only the four-dimensional momentum integral but also the KK sum,
and our goal is to implement these two regulators in such a balanced way that
both five-dimensional Lorentz invariance
and five-dimensional gauge invariance are maintained.
It is this ``balancing'' feature which extends the 't~Hooft-Veltman dimensional regularization
procedure to spacetimes with compactified extra dimensions, and which results
in our name ``extended dimensional regularization'' (EDR).

As we shall see, the EDR procedure will
consist of three separate components:
\begin{itemize}
\item  First, we shift the 4D momentum integral into $d\equiv 4-\epsilon$ spacetime
       dimensions.  
\item  At the same time, we deform the integrand 
        $\Omega_n^{MN...}(\ell_E,\ell^4,x)$ to reflect the fact that our integral
       is now in $d\equiv 4-\epsilon$ dimensions.
       For example, one standard integrand substitution which is familiar from
       traditional dimensional regularization in four dimensions is
       to replace $\ell^\mu \ell^\nu \to \ell^2 g^{\mu\nu}/(4-\epsilon)$ where
       $\ell^2\equiv g_{\mu\nu}\ell^\mu\ell^\nu$.
       However, we now expect there to be a similar deformation for the terms
       in the integrand which depend on the (discrete) fifth component $\ell^4$
       of the momentum.  Deriving the precise
       form of this deformation is the first of our tasks.
       Note that since the introduction of a fifth dimension does not introduce
       any additional Dirac $\gamma$-matrices,
       the usual deformation of the $\gamma$-matrix algebra
        that one must perform for 4D dimensional regularization is unchanged
        for 5D.
\item  Finally, we apply lower and upper cutoffs 
       $\lbrace r_1(\epsilon), r_2(\epsilon)\rbrace$ to our
       KK sum, so that this sum is over the range $r_1(\epsilon)\leq r \leq r_2(\epsilon)$.
       These cutoffs will be functions of $\epsilon$, and deriving the precise relation 
       between $\epsilon$ and these limits is our second task.
\end{itemize}
Indeed, the precise deformation of terms involving $\ell^4$ 
in the integrand, as well as the precise forms of the cutoffs 
$\lbrace r_1(\epsilon), r_2(\epsilon)\rbrace $
as functions of $\epsilon$, will be
determined by the fact that {\it five-dimensional}\/
Lorentz invariance and gauge invariance must be maintained.

Even before imposing five-dimensional gauge invariance,
there are certain simplifications we can make.
First, we know that we must have    
$r_1(\epsilon)\to -\infty$ and
$r_2(\epsilon)\to +\infty$ as $\epsilon\to 0$.
Second, however, just as in Eq.~(\ref{rcut}), we claim that 
$r_{1,2}(\epsilon)$ must actually take the form
\beqn
           r_1(\epsilon) &=& -\Lambda(\epsilon) R + x n\nonumber\\
           r_2(\epsilon) &=& \phantom{-}\Lambda(\epsilon) R + x n
\label{klowup}
\eeqn
in terms of a single as-yet-undetermined function $\Lambda(\epsilon)$. 
In other words, although our summation cutoffs are not symmetric in the $r$-variable,
we claim that they must be symmetric in the $\ell^4$-variable, where $R\ell^4\equiv r-xn$. 
The reason for this is simple.
At first glance, it might appear that
since the four-momentum integrals in dimensional regularization
are over infinite domains, 
there is no difference between integrating over the
internal loop four-momentum $k^\mu$
or the shifted loop four-momentum $\ell^\mu \equiv k^\mu - x p^\mu$,
and we might expect the same
to hold for the KK sums. 
However, integrals which are odd with respect to $\ell$
vanish by convention in dimensional regularization. 
This means that it is the domain of integration for $\ell$ which is symmetric, 
even if it tends to an infinite size. Therefore, higher-dimensional Lorentz invariance
requires that the limits on $\ell^4$ also be the ones which are 
symmetric. 
Indeed, we have verified that any other choice will ultimately lead to 
inconsistencies --- specifically, the sorts of checks that we will perform
at the end of Sect.~4 would not be successful with any other choice.

We can also further refine the form of the deformations
within the integrand 
$\Omega_n^{MN...}(\ell_E,r,x)$ itself.
As mentioned above, we know that terms of the form $\ell^\mu\ell^\nu$
should be replaced by $\ell^2 g^{\mu \nu}/(4-\epsilon)$.
In flat space (which is the only case we consider in this paper), this 
amounts to a deformation for terms $(\ell^{i})^2$ for $i=0,1,2,3$.
Five-dimensional Lorentz invariance therefore requires a
corresponding deformation
for the discrete {\it fifth}\/ component $\ell^4\ell^4$ 
that arises within expressions of the form $\ell^M\ell^N$.
In general, we can parametrize this deformation in the form
\beq
           \ell^4\ell^4 ~\to~ \left[ 1 + \lambda \epsilon + {\cal O}(\epsilon^2)\right] (\ell^4)^2~
\label{kdeformation}
\eeq
where $\lambda$ is an unknown parameter we seek to determine.
As we shall see, determining the deformation to this order in $\epsilon$
will be sufficient for our purposes.
We stress, however, that the deformation in Eq.~(\ref{kdeformation}) is only
appropriate for terms that arise within a Lorentz-covariant expression 
of the form $\ell^M\ell^N$.
By contrast, terms $(\ell^4)^2$ which might arise
from other Lorentz-covariant forms such as $[\ell^2-(\ell^4)^2]g^{MN}$
remain undeformed, in accordance with our expectations from ordinary 
dimensional-regularization in four dimensions.

Given these observations, we 
can then proceed by implementing the variable substitutions described
in Sect.~2.
We thus have
\beq
   L_0^{MN...} ~=~ i\int_{0}^{1}dx \, \sum_{r = -\Lambda R}^{\Lambda R}
            \, \int \frac{d^d\ell_E}{(2\pi)^d} 
                  ~ \Omega_0^{MN...}(\ell_E,r,x) 
\label{Ldim_0}
\eeq
where the zero KK subscript indicates that all external particles are zero
modes, and
\beq
    L_n^{MN...} ~=~ i \int_{0}^{1}d\hat{u}~
         \frac{1}{|n|}\sum_{j = 0}^{|n| - 1}\,
    \sum_{\hat{r} = -\Lambda R + 1}^{\Lambda R}\, 
            \int \frac{d^d\ell_E}{(2\pi)^d} 
                  ~ \Omega_n^{MN...}(\ell_E,\hat r,\hat u, j) 
\label{Ldim_n}
\eeq
where the transformed variables $\hat{u}$ and $\hat{r}$ are defined in
Eqs.~(\ref{u_hat}) and (\ref{r_hat}). 
As discussed above, the KK cutoffs $\Lambda$ are to be viewed as functions
of $\epsilon$. 

Again, we stress that it is remarkable that there will exist solutions
for $\Lambda(\epsilon)$ and $ \lambda$ which can simultaneously preserve
both higher-dimensional Lorentz invariance and higher-dimensional gauge
invariance.  After all, our four-momentum integrals are unrestricted,
while our KK summations are truncated.  Likewise, our four-momenta are
continuous, while our KK momenta are discrete.
Nevertheless, we shall find that the proper solutions for $\Lambda(\epsilon)$
and $\lambda$ will conspire to simultaneously maintain both of these higher-dimensional 
symmetries at the end of any calculation. 

Thus, the complete development of our EDR regulator
now rests on determining two remaining unknowns.  
First, we seek to determine $\Lambda(\epsilon)$ as a function
of $\epsilon$.  Second, we seek to determine the value of the parameter
$\lambda$ in Eq.~(\ref{kdeformation}).

\subsection{Ward-Takahashi identities for KK photons
\label{ward}}

We now demand that our EDR regulator preserve
whatever five-dimensional gauge invariance 
exists prior to compactification.
However, before proceeding further, it is important to determine the extent to 
which the process of compactification, in and of itself,
might break the full five-dimensional
gauge invariance.  In other words, we need to understand the extent to
which five-dimensional gauge invariance can be expected to survive
the process of spacetime compactification.

In this section, we shall address this issue within
the framework of the specific case of five-dimensional QED
compactified on a circle.
Although the usual Ward identities (and indeed the more
general Ward-Takahashi identities) are expected to 
hold for the usual four-dimensional zero-mode photon 
(as a result of the residual {\it four}\/-dimensional gauge invariance), 
we shall demonstrate that {\it analogues of these identities 
actually hold for all of the KK excitations of the photon as well}\/.
In other words, five-dimensional gauge invariance is manifested in
our compactified theory through the existence of a whole tower of Ward(-Takahashi)
identities, one for each KK photon excitation;
compactification
does not break gauge invariance at the level of these
identities.  
As such, these identities can be taken as the 
signature of the original full five-dimensional gauge invariance,
and demanding that these identities continue to hold in our compactified
theory will ultimately enable us to determine 
the  value for the parameter $\lambda$ as well
as the relation between $\Lambda$ and $\epsilon$. 

Let us begin by quickly reviewing the 
usual four-dimensional Ward and Ward-Takahashi identities.
Let $\mathcal{M}(p;q_1,...,q_N;q'_1,...,q'_N)$ represent the
full Fourier-transformed correlation function for some 
QED process with $N$ incoming fermions of
four-momenta $\lbrace q_1,...,q_N\rbrace$, $N$
outgoing fermions with four-momenta $\lbrace q'_1,...,q'_N\rbrace$, 
and an incoming photon $\gamma$ with four-momentum $p$. 
In general, these fermion momenta need not be on-shell,
and we can write $\calM$ in the form
$\mathcal{M} = \epsilon_\mu \mathcal{M}^\mu$
where $\epsilon_\mu$ represents the photon polarization four-vector.
Likewise, let
$\mathcal{M}_0$ represent the full Fourier-transformed correlation
function for the same process 
except without the photon $\gamma$. 
Then the usual four-dimensional Ward-Takahashi identity states that
\beqn
    p_\mu \mathcal{M}^\mu (p;q_1,...,q_N;q'_1,...,q'_N) 
  &=& e\, \sum_i \Biggl[ \mathcal{M}_0 (q_1,...,q_n;q'_1,...,(q'_i - p),...) \nonumber \\
  && ~~~~- \mathcal{M}_0 (q_1,...,(q_i + p),...;q'_1,...,q'_N)\Biggr]~,
\label{ward_t}
\eeqn
where $e$ is the unit of electric charge carried by each fermion. 
Moreover, if we then use the LSZ reduction procedure to obtain
the corresponding amplitude for the corresponding amputated diagrams,
we find that the right side of Eq.~(\ref{ward_t}) does not contribute.
We thus obtain the simpler Ward identity 
\beq
    p_\mu \, \mathcal{M}^\mu (p;q_1,...,q_N;q'_1,...,q'_N) ~=~ 0~
\label{kward}
\eeq
which holds when each of the external momenta 
(including that of the external photon) is on shell.
Of course, the quantity $\calM$ in Eq.~(\ref{kward}) now represents
the amplitude of the corresponding amputated diagram, and the external
momenta are now restricted to be on-shell.

\begin{figure}[ht]
\centerline{
   \epsfxsize 3.7 truein \epsfbox {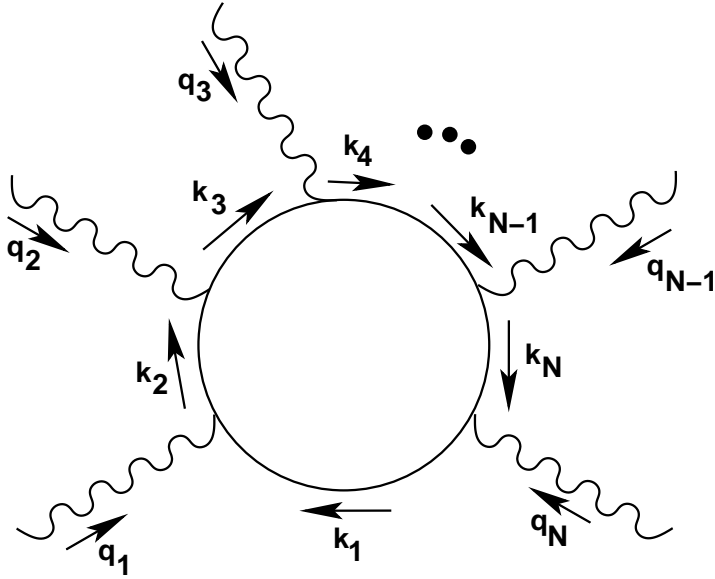} }
\caption{A closed fermion loop with $N$ photon lines, with momentum
     labeling conventions as indicated.
    Summing over all possible insertion locations of an additional
    photon with momentum $p^\mu$ into this diagram produces the amplitude in 
      Eq.~(\protect\ref{4loop}).}
\label{fig3}
\end{figure}

Before we consider whether and how these identities 
can be extended to the case of a compactified higher-dimensional spacetime,
we first review their derivation.
The usual diagrammatic proof of the Ward-Takahashi identity (see, \eg,
any standard reference such as Ref.~\cite{Peskin}) proceeds by realizing
that by summing over each of the diagrams that contribute to $\calM_0$,
and then by summing over all possible ways of inserting an extra external
photon into each of these diagrams, we produce all of the diagrams
contributing to $\calM$.  Thus, we can focus on any individual diagram
contributing to $\calM_0$, and consider all possible ways in which an additional
external photon line can be inserted into such a diagram.
In QED, a photon line can only be inserted onto an already-existing fermion line,
and there are only two possible types of fermion lines such a diagram may contain:  
a closed internal loop (as illustrated in Fig.~\ref{fig3}), 
or a line which ultimately connects an incoming fermion to 
an outgoing fermion.

If the additional photon connects to a fermion line in the former class,
the sum over insertion locations cancels identically upon integrating over the
internal fermion loop momentum.
Specifically, the sum over all insertion points for a photon
of momentum $p^\mu$  into the diagram in Fig.~\ref{fig3} is
proportional to
\beqn
    && e^{N+1}\, \int \frac{d^4 k_1}{(2\pi)^4} ~ {\rm tr}~ \Bigg[ \left( \frac{i}{\not{k}_N
     - m}\right) \gamma^{\lambda_N}\left(\frac{i}{\not{k}_{N - 1} -
     m}\right)\gamma^{\lambda_{N - 1}} \, ...\, \left(\frac{i}{\not{k}_1 -
     m}\right)\gamma^{\lambda_1} \nonumber \\ 
   && ~~~~~~- \left(\frac{i}{\not{k}_N
     + \not{p} - m}\right)\gamma^{\lambda_N}\left(\frac{i}{\not{k}_{N - 1}
     + \not{p} - m}\right)\gamma^{\lambda_{N - 1}} \, ... \, \left(
     \frac{i}{\not{k}_1 + \not{p} - m}\right) \gamma^{\lambda_1} \Bigg]~.~~~~~~~
\label{4loop}
\eeqn
The $\gamma^{\lambda_j}$ factors are from the vertices of the photons
already shown in Fig.~\ref{fig3}, and $m$ is the mass of  the internal fermion
running in the loop. 
[Note that these momenta $k_i$, $q_i$ and the integer $N$
have no relations to the similarly-named 
quantities in Eqs.~(\ref{ward_t}) and (\ref{kward}).]
However, it is easy to see that Eq.~(\ref{4loop}) vanishes.
Rewriting Eq.~(\ref{4loop}) as the difference of two integrals,
we can shift the variable of
integration in the second term from $k_1$ to $k_1 + p$. 
These two integrals thus cancel against each other identically. 
We see, then, that sum over all insertion points of a photon into a
closed loop is zero;  such diagrams do not contribute to right side of
the Ward-Takahashi identity.

By contrast, the right side of Eq.~(\ref{ward_t}) arises from the 
subclass of diagrams in which the additional photon line attaches
to a fermion line that connects an incoming fermion to an outgoing fermion.  
The treatment of such diagrams is standard, and the derivation can be found in Ref.~\cite{Peskin}.
The upshot is that the summation over diagrams contributing to $\calM_0$ then yields Eq.~(\ref{ward_t}).
Although this is only a diagrammatic proof
of the Ward-Takahashi identity, it is  sufficient for our purposes
and can be replaced by a more general path-integral derivation if needed.

We now wish to extend this derivation of the Ward-Takahashi identity 
to the case of five-dimensional 
QED compactified on a circle. 
Our first step will be to repeat this derivation in five
 {\it uncompactified}\/ dimensions.
However, it is immediately clear that there is no change to the basic
result.  Indeed, the entire diagrammatic proof sketched above survives intact, 
and we obtain a five-dimensional Ward-Takahashi identity which is
identical to Eq.~(\ref{ward_t}) except with the replacement 
of Lorentz indices $\mu\to M\equiv (\mu, 4)$ and the understanding
that all momenta are now {\it five}\/-momenta. 
Thus, each five-momentum now contains the usual four-momentum
as well as an additional fifth component.
The same is true, of course, for the external photon momentum $p$.

Given this, our second and final step is to determine the extent to 
which this five-dimensional Ward-Takahashi identity survives the
process of compactification.   Of course, compactification
has the net effect of changing each of these fifth components from
continuous to discrete.
For cases in which the external photon attaches to a fermion line stretching
between incoming and outgoing fermions, 
this discretization of the fifth component has no net effect on the analysis
and our algebraic results survive as before.

However, we must also verify that there are no new features for the cases in
which the external photon attaches to a fermion line which forms a closed
internal loop.  
This case is special because our integral over the internal
loop five-momentum now becomes an integration over the four-dimensional 
loop-momentum components as well as a discrete summation over the fifth component 
(\ie, a summation over the Kaluza-Klein index of the internal fermion).
To be more specific,
we now wish to consider the compactified five-dimensional
analogue of Fig.~\ref{fig3} in which each of the momenta shown represents
a discretized five-momentum, with $k_i\equiv (k_i^\mu, k_i^4)$ 
and $q_i\equiv (q_i^\mu, q_i^4)$ where
$k_i^4\equiv r_i/R$ and $q_i^4\equiv s_i/R$ 
for some integers $r_i,s_i\in \IZ$.
If our external photon has five-momentum $p\equiv (p^\mu, n/R)$,
the sum over insertion locations for this external photon now leads to
the compactified five-dimensional amplitude
\beqn
    && e^{N+1}\, \sum_{r\in \IZ}\,
    \int \frac{d^4 k_1}{(2\pi)^4} ~ {\rm tr}~ \Bigg[ \left( \frac{i}{\not{k}_N
     - m}\right) \gamma^{\lambda_N}\left(\frac{i}{\not{k}_{N - 1} -
     m}\right)\gamma^{\lambda_{N - 1}} ... \left(\frac{i}{\not{k}_1 -
     m}\right)\gamma^{\lambda_1} \nonumber \\ 
   && ~~~~~~- \left(\frac{i}{\not{k}_N
     + \not{p} - m}\right)\gamma^{\lambda_N}\left(\frac{i}{\not{k}_{N - 1}
     + \not{p} - m}\right)\gamma^{\lambda_{N - 1}} ... \left(
     \frac{i}{\not{k}_1 + \not{p} - m}\right) \gamma^{\lambda_1} \Bigg]
     \nonumber \\
\label{k5version}
\eeqn
where quantities such as $\not{k}$ are now understood to represent
five-dimensional contractions, \ie, 
$\not{k}\equiv k_M \gamma^M \equiv k_\mu \gamma^\mu - (r/R) \tilde\gamma^4$
where $\tilde \gamma^4\equiv i\gamma^5 = \gamma^0 \gamma^1 \gamma^2 \gamma^3$. 
Just as with Eq.~(\ref{4loop}), 
we can once again separate these terms into distinct integrations/summations
and recognize that the second term is the same as the first term except
for the algebraic replacements $k_i^\mu \to k_i^\mu  + p^\mu$
and $r_i\to r_i+n$.
The first of these replacements has no net effect because the four-momentum 
integration in Eq.~(\ref{k5version}) has infinite range;  indeed, this
range remains infinite even when the integrand is regulated through 
4D dimensional regulation.
However, the situation with the second replacement is slightly more subtle.
Of course, the shift $r_i\to r_i+n$ does not disturb the form of our KK summation 
because each integer $r_i$ in the summation range is merely being shifted by 
another integer $n$.
However, in this case the summation range is not infinite, since
there is an implicit cutoff.
It is therefore only as this cutoff is removed at the end of the calculation
that the replacement $r_i\to r_i+n$ has no net effect on the KK summation,
and Eq.~(\ref{k5version}) holds.
Of course, for the special $n=0$ case (corresponding to a zero-mode external photon),
this last issue does not arise, and the KK summation is unaltered regardless
of the value of the cutoff.

Putting the pieces together, then, we obtain
a Ward-Takahashi identity which is suitable for five-dimensional 
spacetimes with a single compactified dimension:
\beqn
    p_M \, \mathcal{M}^M (p;k_1,...,k_N;q_1,...,q_N) 
  &=& e\, \sum_i \Biggl[ \mathcal{M}_0 (k_1,...,k_n;q_1,...,(q_i - p),...) \nonumber \\
  && ~~~~- \mathcal{M}_0 (k_1,...,(k_i + p),...;q_1,...,q_N)\Biggr]~.~~~~
\label{5ward_t}
\eeqn
Here $M$ is the five-dimensional Lorentz index, and all momenta are understood
to be {\it five}-momenta.
As usual, this identity holds in the presence of a suitable regulator. 
In the special case of a zero-mode external photon, this identity holds exactly;
by contrast, for all other cases, this identity 
holds {\it up to terms which vanish as the regulator is removed}. 
The identity in Eq.~(\ref{5ward_t}) is quite powerful, however:  it implies not only 
that our ordinary (zero-mode) photon satisfies the Ward-Takahashi
identity (as we might have always expected), but also that {\it each of our 
excited KK photons satisfies a Ward-Takahashi identity as well}\/.  In this sense,
our original five-dimensional gauge invariance has survived
the process of compactification --- even though our original five-dimensional
Lorentz invariance is broken.  

Given this result, we can then generate a corresponding five-dimensional 
Ward identity in the usual way.  In general, Ward identities follow from the
Ward-Takahashi identities through LSZ reductions, but we do not really
require the full LSZ machinery.
The critical observation is that the two sides of Eq.~(\ref{5ward_t}), just like 
the two sides of its four-dimensional
version Eq.~(\ref{ward_t}), have differing pole structures in momentum space:
the left sides of these equations have $2N+1$ poles, while
the right sides of these equations have $2N$ poles.
Nothing pertaining to the dimensionality of the spacetime or
the process of compactification reconciles this mismatch in the pole structure.
Consequently, passing to the amplitudes of the corresponding {\it amputated}\/ diagrams
and placing our external particles on shell,
we find that the right sides of these equations cannot contribute, and
thus we obtain a five-dimensional Ward identity which holds for each KK photon:
\beq
    p_M \, \mathcal{M}^M (p;k_1,...,k_N;q_1,...,q_N)~=~0~.
\label{k5DWard}
\eeq
As with Eq.~(\ref{5ward_t}), it is understood that this is an exact
relation which holds 
for zero-mode external photons
in the presence of a regulator; 
for excited KK photons, by contrast,
this relation holds 
up to terms which vanish as the regulator is removed. 
However, this will be sufficient for our purposes.

Finally, note that unlike the 
Ward-Takahashi identities
in Eq.~(\ref{5ward_t}),
the Ward identities in Eq.~(\ref{k5DWard}) hold only when
the external photon is on-shell.
However, in the special case of amplitudes with no external fermions,
the right side of Eq.~(\ref{5ward_t}) vanishes identically.
In such cases, we expect the Ward identity in Eq.~(\ref{k5DWard}) to 
hold regardless of whether the external photon momentum is on-shell
or off-shell.

One important special case that we will shortly consider is the case 
of diagrams with two external photons and no external fermions ---
 \ie, a five-dimensional vacuum polarization diagram.
By momentum conservation, the five-momentum $p^M=(p^\mu,n/R)$ of the incoming
photon will be equal to the five-momentum of the outgoing photon.
In this case, our amplitude $\calM^{MN}$ will have two five-dimensional
Lorentz indices, and our Ward identities
take the form
\beq
            p_M \, \calM^{MN} ~=~ p_N \, \calM^{MN}~=~ 0~.
\label{kward1}
\eeq
Expanded out, these identities imply
\beq
           p_\mu \, \calM^{\mu\nu} ~=~ {n\over R} \, \calM^{4\nu}
          ~~~~~ {\rm and} ~~~~~ 
           p_\nu \, \calM^{\mu\nu} ~=~ {n\over R} \, \calM^{\mu 4}
\label{kward2}
\eeq
as well as 
\beq
           p_\mu \, \calM^{\mu 4} ~=~ {n\over R} \, \calM^{44}
          ~~~~~ {\rm and} ~~~~~ 
           p_\nu \, \calM^{4 \nu} ~=~ {n\over R} \, \calM^{44}~.
\label{kward3}
\eeq
Combining these two results, we thus obtain the relation
\beq
           p_\mu p_\nu \, \calM^{\mu\nu} ~=~ \left( {n\over R}\right)^2 \, \calM^{44}~.
\label{kward4}
\eeq

Of course, our derivation of these identities has been purely diagrammatic
and restricted to the special case of five-dimensional QED compactified on a circle.
Despite these limitations, the arguments of this section should easily  
generalize to the case of multiple extra dimensions
compactified on square tori. 
Moreover, we expect identities like these to hold for even more
general spacetimes and compactifications.
After all, Ward(-Takahashi) identities are merely expressions
of Noether's theorem (and resulting Schwinger-Dyson equations) 
applied to gauge symmetries.
As such, they can generally be proven using 
path-integral techniques which should survive 
compactification as long as 
no spacetime boundary is introduced (to produce new surface terms).
Thus, we expect a Ward identity of this type to emerge 
whenever our higher-dimensional Lagrangian exhibits a gauge symmetry
and the spacetime is compactified on a smooth manifold.

Needless to say, the situation can be significantly different for 
compactifications on orbifolds.  The presence of fixed 
points (or fixed lines/planes, {\it etc.}\/)
can give rise to surface terms (such as brane kinetic terms)
which render the would-be Ward identities  
invalid for all but the usual four-dimensional Ward identity on the brane.
Moreover, even for compactifications on manifolds, we stress that
the corresponding Ward identities may not always take a recognizable form.
Implicit in our derivation above was the assumption that the  
Kaluza-Klein eigenfunctions coincide with momentum eigenfunctions.
While this is true for compactifications on square tori,
this will not be true in general:  for example, compactification
on a sphere produces Legendre polynomials which have no interpretations
in terms of individual plane waves. 
Since our Ward identities are usually written in terms of momentum-space
wavefunctions, such compactifications can lead to Ward identities
involving many non-trivial interactions between individual plane-wave
modes.

Finally,
we remind the reader that not every regulator will 
respect these identities:  certain UV divergences can
spoil the argument we made about insertions into a KK fermion
loop.  For example, some regulators (\eg, the hard cutoff) are known to violate
these identities in four dimensions.
Thus, only certain regulators will respect these five-dimensional 
Ward(-Takahashi) identities, and it is the goal of this section
to determine for which regulators this is the case.

\subsection{Imposing the Ward-Takahashi identities for KK photons
\label{sumstraint}}

We now impose our higher-dimensional
Ward-Takahashi identities in order to derive a relationship 
between the dimensional-regularization parameter $\epsilon$ and 
the summation cutoff $\Lambda$ introduced in Sect.~\ref{prelim}. 
We shall also determine the precise value for $\lambda$ introduced
in Eq.~(\ref{kdeformation}).

To do this, we consider the special case of
Fig.~\ref{fig1} in which the external particles are on-shell KK
photons and the particles running in the loop are a tower
of KK fermions with bare mass $M$ (so that the tree-level squared
mass of the $r^{\rm th}$ excitation is given by $r^2/R^2 + M^2$).
Such a diagram is indeed nothing but a five-dimensional
vacuum polarization diagram with two Lorentz indices $(M,N)$ corresponding
to the external photons, and 
this is precisely the sort of diagram for which we expect
the higher-dimensional Ward identities given in 
Eqs.~(\ref{kward1}) though (\ref{kward4}) to hold.

Prior to regularization, the different components of the vacuum
polarization amplitude take the form
\beq
    L_n^{M N} ~=~ -4e^2 
          \int_{0}^{1}dx\, \sum_r\, \int \frac{d^4 \ell}{(2\pi)^4} 
           \left[ \frac{1}{\ell^2 - {\ell^4}^2 - \mathcal{M}^2 (x)} \right]^2\,
           \Omega^{MN}_n
\label{origtensor}
\eeq
where
\beqn 
\Omega^{\mu\nu}_n &=& 2\ell^{\mu}\ell^{\nu} + 2x(x - 1)p^{\mu}p^{\nu} \nonumber\\
        && ~~~~~~ +  g^{\mu \nu}
    \left[-\ell^2 + {\ell^4}^2 + (2x - 1)(n/R)\ell^4 - 
              \mathcal{M}^2 (x) + 2M^2 \right]\nonumber\\
\Omega^{\mu 4}_n &=&  p^{\mu} \left[(2x - 1)(n/R)\ell^4 + 2x(x - 1)n/R\right]\nonumber\\
\Omega^{4 4}_n &=&   \ell^{2} + {\ell^4}^2 + (2x - 1)(n/R)\ell^4 + 
             2x(x - 1)n^2 /R^2 + \mathcal{M}^2 (x) - 2M^2~.~~~~ 
\label{mu_nu_int}
\eeqn
Here $n$ is the mode number of the external photon,
and in writing these expressions, we have continued to use the notation
and conventions listed at the beginning of Sect.~2.
The procedure outlined in Sect.~\ref{prelim}
then demands that we 
regularize four-momentum
integrals by taking their dimensionality to be $d = 4 - \epsilon$, 
truncate KK sums according to Eq.~(\ref{klowup}),
and also deform
the integrands according to Eq.~(\ref{kdeformation}).
After performing the momentum loop integrations, we then
find that these components take the form
\beq
    L_n^{M N} ~=~ - { i e^2 \over 4\pi^2}\, R^\epsilon \,\int_0^1 dx\, 
        \sum_{r=-\Lambda(\epsilon)R+xn}^{\Lambda(\epsilon)R+xn} \, f^{MN}_n
\label{kdform1}
\eeq
where
\beqn
   f^{\mu\nu}_n &=& 
     \left\lbrace \left[ (2x - 1)(n/R)\ell^4 + 2M^2 - 2\mathcal{M}^2 (x)\right]
            g^{\mu \nu} + 2x(x - 1)p^{\mu}p^{\nu} \right\rbrace \, W~~~~\nonumber\\
   f^{\mu 4}_n  &=& 
         p^{\mu}
          \left[(2x - 1)\ell^4 + 2x(x - 1)(n/R)\right]\, W~\nonumber\\ 
   f^{4 4}_n  &=& \left[3{\ell^4}^2 + (2x - 1)(n/R)\ell^4 + 2x(x - 1)(n/R)^2 + 
           3\mathcal{M}^2 (x) - 2M^2 \right] 
           \, W \nonumber\\
            && ~~~~+ (1 - 2\lambda){\ell^4}^2 + \mathcal{M}^2 (x)
\label{kdform2}
\eeqn
with
\beq
 W ~\equiv~ \frac{2}{\epsilon} - \gamma + \log(4\pi) -
    \log[(\ell^4 R)^2 + (\mathcal{M}(x)R)^2] +  \mathcal{O}(\epsilon)~.
\label{kdform3}
\eeq
Here $\gamma$ is the Euler-Mascheroni constant.
Note that the 
KK summation in Eq.~(\ref{kdform1})
does not necessarily force the terms which are
linear with respect to $\ell^4$ to vanish.  This is an important
distinction from the case in which $\ell^4$ is a continuous variable.

Given these expressions for the vacuum polarization diagrams, 
we now demand that they respect the Ward
identities~(\ref{kward2}) and (\ref{kward3})
for the KK photon modes.
First, we immediately observe from the above results that 
\beq
     p_\mu \, f^{\mu\nu}_n ~=~ \left( {n\over R} \right)\, f^{4\nu}_n~.
\label{wardintegrands}
\eeq
Thus, we find that the full Ward identity in Eq.~(\ref{kward2}) for the
 {\it amplitudes}\/ $L_n^{MN}$ is satisfied identically as the result of 
a Ward identity for
the {\it integrands}\/ $f^{MN}_n$ for all external KK photon mode numbers $n$.   
This implies that the Ward identity in Eq.~(\ref{kward2}) holds 
regardless of whether the external KK photon is on-shell or off-shell,
and regardless of how $\epsilon$ and $\Lambda$ are related in the 
internal KK sum in Eq.~(\ref{kdform1}).
Moreover, because this amplitude contains no external fermions,
the fact that the Ward identities hold when the external photon
momenta are off-shell implies that the full Ward-Takahashi identities
hold as well. 
Thus, while Eq.~(\ref{wardintegrands}) 
is an important self-consistency check on our approach,
it does not yield any new information which helps us determine $\Lambda(\epsilon)$
or $\lambda$.

The situation, however, is different for the Ward identity in Eq.~(\ref{kward3}).
Examining the integrands, we find that
\beqn
   p_\mu f^{\mu 4} - \left( {n\over R}\right) f^{44}  &=& 
         \left[ -\left({n\over R}\right)\left(3 {\ell^4}^2 + \calM^2(x) \right)
         + (2x-1) \ell^4 \left( p^2-{n^2\over R^2}\right) \right]~W
\nonumber\\
    && ~~~~~~ 
      -\left( {n\over R}\right) \left[ (1-2\lambda) {\ell^4}^2 + \calM^2(x) \right]~. 
\label{kform5}
\eeqn 
Note that this vanishes identically when our external photon is the zero-mode
photon ({\it i.e.}\/, $n=0$) and is on-shell.  
Thus, we find that the Ward identity in Eq.~(\ref{kward3}) also holds automatically
for zero-mode photons, as we expect.  
Moreover,
even when the external zero-mode photon
is not on-shell, the Ward(-Takahashi) identity continues to hold because
the non-zero factor $(p^2-n^2/R^2)$ in Eq.~(\ref{kform5}) 
comes multiplied by a single power of $\ell^4$, which vanishes over the 
symmetric $r$-summation
in Eq.~(\ref{kdform1}). 
Together, this is nothing but the preservation of
four-dimensional gauge invariance, which once again occurs 
regardless of the precise relations between $\Lambda$, $\lambda$, or $\epsilon$.   
  
By contrast, in order to preserve {\it five}\/-dimensional gauge invariance,
we require that the Ward(-Takahashi) identities in Eq.~(\ref{kward3}) hold for {\it all}\/
KK photons --- {\it i.e.}\/, for {\it all}\/ values of $n$.
We must therefore concentrate on the cases when $n\not =0$,
and determine a value for $\lambda$ and a relation between the KK summation cutoff
$\Lambda$ and $\epsilon$ such that Eq.~(\ref{kward3}) holds. 
At first glance, our main complication is that 
our cutoffs $\Lambda$ appear only in the KK summation limits.
However, since $n\not=0$, we can utilize the variable-transformation methods we
developed in Sect.~2.
Specifically, following the steps outlined in  
Sect.~2, we change variables from 
$x$ to $\hat u$ defined in Eq.~(\ref{u_hat})
and from
$r$ to $\hat r$ defined in Eq.~(\ref{r_hat}),
and then drop the hats from $\hat u$ and $\hat r$.
This amounts to the algebraic substitution
$x\to (u+j)/|n|$, and we shall define
$y\equiv (u+j)/|n|$.
Following Eq.~(\ref{Ldim_n}),
we can then write
\beq
        p_{\mu} L^{\mu 4} - \frac{n}{R} L^{4 4}  ~=~
       {ie^2 \over 4\pi^2}\,
       {{\rm sign}(n)R^\epsilon\over R}\,
          \sum_{r = -\Lambda R + 1}^{\Lambda R}\,
        \frac{1}{|n|}\, 
         \sum_{j = 0}^{|n| - 1}\, \int_{0}^{1}du ~  f_n
\label{4_comp2}
\eeq
where the integrand $f_n$ is the variable-shifted
version of Eq.~(\ref{kform5}), {\it i.e.}\/, 
\beqn
      f_n  &=&
         \left[ |n|\left({3(r-u)^2\over R^2}  + \calM^2(y) \right)
         + (1-2y) (r-u) \left( p^2-{n^2\over R^2}\right) \right]~W
\nonumber\\
    && ~~~~~~ 
      + |n| \left[ (1-2\lambda) {(r-u)^2\over R^2}  + \calM^2(y) \right]~. 
\label{kform6}
\eeqn 
Here $W$ continues to represent the quantity in Eq.~(\ref{kdform3}),
now written with the algebraic substitutions $(R\ell^4)^2\to (r-u)^2$
and $\calM^2(x)\to \calM^2(y)$. 

It is not immediately clear which relationships between $\Lambda$, $\lambda$,
and $\epsilon$ 
would force the expression in Eq.~(\ref{4_comp2}) to vanish
as $\Lambda\to \infty$ (or as $\epsilon\to 0$),
or whether such a relation even exists. 
However, we may consider the special case in which 
the external KK photons are on-shell.
In other words, we can restrict our attention to the Ward identities
rather than the full Ward-Takahashi identities.
Once we determine the appropriate relationships between 
$\Lambda$, $\lambda$, and $\epsilon$ for the purposes of maintaining the
Ward identities, we can then verify
that the full Ward-Takahashi identities hold as well.

When the external KK photons are on-shell, $p^2 - n^2 /R^2 = 0$ 
and $\mathcal{M}^2 (y) = M^2$. 
Our integrand is also independent of $j$, which 
enables us to explicitly perform the $j$-summation
in Eq.~(\ref{4_comp2}) and soak up the overall factor of $|n|$.
We then see that Eq.~(\ref{4_comp2}) 
is given by
\beqn
    && p_{\mu} L^{\mu 4} - \left({n\over R}\right) L^{4 4} \nonumber\\
       &&~~~~~~~=~ 
          {ie^2 n R^\epsilon \over 4\pi^2 R}
          \sum_{r = -\Lambda R + 1}^{\Lambda R}
         \int_{0}^{1}du  
          \Biggl[ \left({3(r-u)^2\over R^2}  + M^2 \right) \, W 
              +   (1-2\lambda) {(r-u)^2\over R^2}  + M^2 \Biggr]\nonumber\\
       &&~~~~~~~=~ 
  {ie^2 nR^\epsilon \over 4\pi^2 R }\,
          \sum_{r' = -\Lambda R }^{\Lambda R-1}\,
         \int_{r'}^{r'+1}dw \, 
          \Biggl[ \left({3w^2\over R^2}  + M^2 \right) \, W 
      +   (1-2\lambda) {w^2\over R^2}  + M^2 \Biggr]~\nonumber\\
       &&~~~~~~~=~ 
  {ie^2 nR^\epsilon \over 4\pi^2 R}\,
         \int_{-\Lambda R}^{\Lambda R}dw \, 
          \Biggl[ \left({3w^2\over R^2}  + M^2 \right) \, W 
      +   (1-2\lambda) {w^2\over R^2}  + M^2 \Biggr]~\nonumber\\
       &&~~~~~~~=~ 
  {ie^2 nR^\epsilon \over 4\pi^2 R}\,
  \Biggl\lbrace    {2\tilde \Lambda^3\over R^2} 
        \left[ 1+c - {2\lambda \over 3} - \log(\tilde \Lambda^2 + M^2 R^2)\right]\nonumber\\
  &&~~~~~~~~~~~~~~~~~~~~~~~~~~~~~~~~~~~~~ + 2\tilde \Lambda M^2 
          \left[ 1+c - \log(\tilde \Lambda^2 + M^2 R^2)\right]\Biggr\rbrace~.
\label{kform7}
\eeqn 
Note that the second equality above follows from defining 
$w\equiv u-r$ and $r'= -r$, and the third follows from explicitly performing
the truncated KK sum.
The fourth equality is obtained by 
substituting $ W = 2/\epsilon - \gamma + \log(4\pi) -
    \log[w^2 + (MR)^2] +  \mathcal{O}(\epsilon)$
and explicitly evaluating the $w$-integral.
Finally, in writing the final line, we have defined $\tilde \Lambda\equiv \Lambda R$
and $c\equiv 2/\epsilon -\gamma +\log(4\pi)$.

Given these results, we see that there are many different ways in which this final
expression can be made to vanish as $\tilde \Lambda\to\infty$, as required by
our Ward identity for excited KK photons. 
One possibility, for example, is to demand that $\Lambda$ and $\epsilon$
be related to each other such that $1+c= \log(\tilde \Lambda^2 + M^2 R^2)$
up to terms which vanish more strongly than $1/\tilde \Lambda^3$ as
$\tilde \Lambda\to\infty$.
If we additionally take $\lambda=0$, then both of the terms in the final
expression in Eq.~(\ref{kform7}) will vanish as $\tilde\Lambda\to \infty$ (or
as $\epsilon\to 0$).
However, 
such relations are not suitable for a bona-fide regulator because they
depend on $M$.  They thus depend on the particular
fermions in the theory, and are not theory-independent.

It turns out that there is only one possible $M$-independent regulator
which does the job.
For large $\Lambda$, we can write 
$\log(\tilde \Lambda^2+M^2R^2)\approx 2\log(\tilde \Lambda) +(MR/\tilde \Lambda)^2$,
whereupon Eq.~(\ref{kform7})
takes the form
\beqn
    p_{\mu} L^{\mu 4} - \left({n\over R}\right) L^{4 4} &=& 
          {ie^2 nR^\epsilon \over 4\pi^2 R}\,
  \Biggl[    {2\tilde \Lambda^3\over R^2} 
        \left( 1+c - {2\lambda \over 3} - 2\log\tilde \Lambda\right)\nonumber\\
  &&~~~~~~~~~~~~~~~~ + 2\tilde \Lambda M^2 
          \left( c - 2\log\tilde \Lambda \right) + {\cal O}(MR/\tilde \Lambda)\Biggr]~.~~~
\label{therdem}
\eeqn
We therefore demand that $c=2\log \tilde \Lambda$ up to terms which
vanish faster than $1/\tilde \Lambda^3$ as $\tilde\Lambda\to 0$,
and we likewise choose $\lambda=3/2$.
These choices guarantee that $p_\mu L^{\mu 4}-(n/R)L^{44}\to 0$
as $\epsilon\to 0$, \ie,  as $\Lambda\to \infty$.

Thus, to summarize,
we conclude that 
the proper relationship between 
$\Lambda$ and $\epsilon$ is given by
\beq
     {2\over\epsilon} - \gamma + \log(4\pi) + 
       \calO(\epsilon) ~=~ 2\,\log(\Lambda R) + \delta
\label{eps_lam}
\eeq
where $\delta \to 0$ as $\Lambda\to\infty$.
[For example, for the expression in Eq.~(\ref{therdem}), we know that
$\delta \Lambda^3\to 0$ as $\Lambda\to \infty$.]
We shall discuss the role played by $\delta$ below.
We also conclude that 
\beq
            \lambda ~=~ 3/2~.
\label{kreg1}
\eeq

Eqs.~(\ref{eps_lam}) and (\ref{kreg1}) 
are the relations between
$\Lambda$, $\lambda$, and $\epsilon$ which preserve higher-dimensional gauge invariance
as well as higher-dimensional Lorentz invariance.
As such, these relations therefore define our extended dimensional-regularization
(EDR) procedure.  
Moreover, as we shall see, these relations are {\it universal}\/
(as demanded by our criterion of theory-independence): 
as we shall soon discuss, they apply for any loop diagram in any theory with a
circular extra dimension, even though we derived them via a study 
of five-dimensional QED.  

Finally, although we have shown above that
these relations are sufficient
to satisfy the Ward identities for all KK photons, we have also verified through an explicit
calculation that they actually satisfy the full Ward-Takahashi identities for KK photons
as well.
In other words, the Ward identities 
are satisfied regardless of whether the external photon momenta are on-shell or off-shell.
 
We should also emphasize an important point.  Clearly, our EDR regulator
should be applicable for all values of the compactification radius $R$. 
As such, the EDR regulator should be applicable even in the $R\to\infty$ limit
in which flat five-dimensional Minkowski space is restored and our KK sum
becomes an integral.
However, even in this limit, our EDR regulator does {\it not}\/ reduce
to ordinary 't~Hooft-Veltman 5D dimensional regularizaton.
This is because we are continuing to treat the resulting five-dimensional
momentum integral in an asymmetric way, even in the $R\to\infty$ limit,
using 4D dimensional regularization
for the large spacetime dimensions and a hard cutoff for the extra spacetime
dimension.  Thus, while we continue to have a self-consistent regulator
even in the $R\to\infty$ limit, this is not the flat five-dimensional
version of the ordinary 't~Hooft-Veltman regulator.
Note that this situation was entirely different for
our extended hard-cutoff regulator in Sect.~2.  In that case,
the $R\to\infty$ limit does reproduce an ordinary five-dimensional hard cutoff.

Another example of this difference between the $R\to\infty$ limit of the EDR
procedure and the ordinary 5D 't~Hooft-Veltman dimensional regularization procedure
is the fact that EDR involves a deformation of the 
four-momentum components of the form $\ell^\mu \ell^\nu\to \ell^2 g^{\mu \nu}/(4-\epsilon)$,
but a deformation of the extra fifth component of the form
in Eq.~(\ref{kdeformation}) with $\lambda=3/2$.
These deformations are intrinsically different, and remain
so even in the $R\to\infty$ limit;  indeed, neither of these deformations
is what would be encountered in 5D 't~Hooft-Veltman dimensional regularizaton.  
These inequivalent deformations in some sense compensate for the inequivalent regularizations
applied to the four-momenta and the KK momenta,
and are precisely what are required in order to 
maintain the Ward-Takahashi identities.  Moreover, as we shall discuss below,
this is also necessary for the maintenance of five-dimensional Lorentz invariance
for all values of $R$.

Despite these differences,
the overall form of the relation~(\ref{eps_lam}) is expected
at a certain intuitive level. 
We know, for example, that the $1/\epsilon$ pole in ordinary 4D dimensional
regulation corresponds to a logarithmic divergence, and a logarithmic
divergence manifests itself as the logarithm of a cutoff $\Lambda$. 
Thus, a relation of the form in Eq.~(\ref{eps_lam}), which relates 
$1/\epsilon$ to $\log(\Lambda)$, is to be expected.  What is non-trivial, by contrast, is that this relation
also preserves five-dimensional {\it gauge}\/ invariance, as expressed
through the preservation of the Ward identities.    
This, of course, was the objective of our entire analysis.


\subsection{Loose ends
\label{sumstraintexact}}

Thus far, our development of the EDR regulator has led us to the conditions
in Eqs.~(\ref{eps_lam}) and (\ref{kreg1}).  However, there are a number
of issues which we have not yet addressed:
\begin{itemize}
\item  We have not yet demonstrated that these conditions preserve
         higher-dimensional Lorentz invariance.
\item  We have not yet demonstrated that these conditions
         are {\it universal}\/ --- \ie, that they suitably regulate
         the divergences that might appear in any potential 
          diagram in a gauge-invariant five-dimensional theory compactified on a circle,
         so that {\it all}\/ possible amplitudes satisfy appropriate Ward-Takahashi
        identities.
\item  And finally, we have not yet discussed the significance of the quantity $\delta$
         which appears in Eq.~(\ref{eps_lam}).
\end{itemize}

All of these issues must be addressed before we can claim 
to have a bona-fide regulator for five-dimensional theories compactified
on a circle.
The purpose of this section is to address each of these issues,
one at a time.

\subsubsection{Higher-dimensional Lorentz invariance}

We begin by considering the issue of higher-dimensional Lorentz invariance.

It is, of course, unavoidable that reducing the dimensionality of our
uncompactified spacetime from four dimensions to $D\equiv 4-\epsilon$ dimensions
breaks higher-dimensional Lorentiz invariance, 
since this dimensional-alteration process cannot regularize discrete KK sums.
Therefore, the best one can do in a dimensional-regularization setup is
to restore the higher-dimensional Lorentz symmetry at the end of a calculation,
just as we restore the Ward identities (and more generally, the Ward-Takahashi
identities) in the $\Lambda\to \infty$ limit.
However, we already know that our extended hard-cutoff (EHC) regularization procedure
in Sect.~3 preserves five-dimensional Lorentz invariance, by construction.
Therefore, within the context of a five-dimensional theory without 
gauge invariance, if we can demonstrate that our 
EHC and EDR procedures
lead to identical results after the cutoffs
are removed, we will have demonstrated that our extended dimensional-regularization
procedure preserves higher-dimensional Lorentz invariance.
Fortunately, we have done this calculation within the context of the effective field
theories of KK modes discussed in Ref.~\cite{paper2}, and the results are positive.

Moreover, even within the calculation we have done in Sect.~3.3, 
it is straightforward to verify that five-dimensional Lorentz 
invariance is preserved.
Recall that we began with a vacuum-polarization 
amplitude in Eq.~(\ref{origtensor}) which
 {\it a priori}\/ transforms as a five-dimensional Lorentz tensor.
However, after we impose our regulator, this expression took the     
form in Eq.~(\ref{kdform1})
where the integrands for the different Lorentz components 
are given in Eq.~(\ref{kdform2}).
Clearly, the forms of these different Lorentz components are quite
different, and it seems that higher-dimensional Lorentz invariance is broken.  
However, if we take the $R\to \infty$ limit, the KK sum in Eq.~(\ref{kdform1})
becomes an integral.  Imposing the relations 
in Eqs.~(\ref{eps_lam}) and (\ref{kreg1}) and assuming that ${\cal M}^2(x)\geq 0$,
we then find that these different components all collapse into the
single form
\beqn 
   L^{M N} &=& 
    -\frac{ie_{5}^2}{8\pi^3}\int_{0}^{1}dx  ~ 2x(1 - x)\,
  \left\lbrace
     [p^2 - (p^4)^2] g^{M N} - p^M p^N \right\rbrace ~W'~
\label{tens}
\eeqn
where 
$e_5\equiv \sqrt{2\pi R}\,e$ is the 5D gauge coupling 
and where
\beq 
      W'~=~ 4\Lambda - 2\pi \sqrt{\mathcal{M}^2 (x)} + \mathcal{O}(m^2 /\Lambda)~.
\eeq
Likewise, similar expressions can be derived for the case with ${\cal M}^2(x)<0$. 
Clearly, the expression in Eq.~(\ref{tens}) transforms as a higher-dimensional
Lorentz tensor. 
We note that this happens only if we impose the relations
in Eq.~(\ref{eps_lam}) and (\ref{kreg1}).

We shall present further explicit evidence of the preservation of five-dimensional Lorentz
invariance in Sect.~\ref{need}.

\subsubsection{Universality}

In this section, we discuss the question of {\it universality}\/ ---
\ie, whether our EDR regulator can suitably regulate
the divergences that might appear in any potential 
one-loop diagram in a gauge-invariant five-dimensional theory 
compactified on a circle.

Thus far, we have only demonstrated that EDR
preserves the higher-dimensional Ward-Takahashi identities for
vacuum polarization diagrams with two external KK photons.  However,
our regulator should respect higher-dimensional gauge symmetry in
general.  This only can happen if our extended dimensional-regularization 
procedure preserves KK Ward identities and Ward-Takahashi identities 
for {\it arbitrary}\/ QED processes in higher dimensions.

Even though there are an infinite number of possible amplitudes in
QED, it is sufficient for our regulator to preserve KK Ward-Takahashi identities
for loop diagrams of the type shown in Fig.~\ref{fig3}, with no external
fermions. 
This is because a divergence from
this type of diagram is the only effect which has the potential to
spoil the proof of the Ward-Takahashi identity that we outlined in
Sect.~3.2.  
Furthermore, power counting in 5D implies that diagrams
with six or more external KK photons should be finite.  
Hence, we only need to check that 
the Ward-Takahashi identities hold for amplitudes with at most five external photons
and no external fermions. 
Note that for such amplitudes, the Ward-Takahashi identity reduces to the
same form as the Ward identity, except that the external photons need
not be on-shell. 

We can therefore consider the cases with $0\leq N \leq 5$ external photons
individually.  Just as elsewhere in this paper, we restrict our attention
to one-loop diagrams. 
\begin{itemize}
\item  $N=0$:  Diagrams of this form with no external photons are 
           mere vacuum bubbles which never contribute to physical amplitudes.
\item  $N=1,3,5$:  In these cases, our amplitudes have odd numbers of external
                photons and vanish as a consequence of Furry's theorem.
               Note that Furry's theorem is itself a direct consequence   
               of charge-conjugation symmetry, and does not rely
               on gauge invariance {\it per se}\/.
               Since our regulator respects charge-conjugation invariance, 
               the KK Ward-Takahashi identities 
              are thus trivially satisfied in each of these cases.  
\item $N=2$:   This is the case we already examined, and we have already shown 
                that our dimensional-regularization procedure
                  respects KK Ward-Takahashi identities for such vacuum
                polarization diagrams.
\end{itemize}

Given these conclusions, it only remains to check that our regulator
preserves the Ward-Takahashi identities in the $N=4$ case, \ie, for 
``box'' diagrams of the type shown in Fig.~\ref{fig3} with four external KK photons.  

Of course, if gauge invariance is truly maintained,
then power counting actually over-estimates the degree of
divergence in each diagram.   This is because gauge invariance generally removes several
powers of divergence from each diagram. 
For example, we have already seen that 
gauge invariance forces the
vacuum polarization diagrams 
to diverge linearly in the summation cutoff $\Lambda$ 
rather than cubically.
In general, inserting extra external photons will also decrease the degree
of divergence.
Therefore, if we can show that the $N=4$ box diagram is actually finite,
then our demonstration of universality is complete.

Evaluating the box diagram is a rather complicated undertaking, 
even in four dimensions~\cite{boxpapers}.
Therefore, rather than providing a direct evaluation in five dimensions,
we shall instead provide an indirect argument that this diagram is indeed finite.
Our argument proceeds as follows.
Let us first consider the $R\to \infty$ limit in which our extra
dimension is completely uncompactified.
In this case, we know that the ordinary 't~Hooft-Veltman 5D dimensional 
regularization procedure~\cite{dimreg} 
provides a valid regulator which preserves the Ward-Takahashi identities.
Given that the Ward-Takahashi identities are satisfied for this regulator,
it can be shown that our 5D box amplitude is finite;  this will be demonstrated
explicitly below.
Thus, we conclude that the box amplitude is finite in the $R\to\infty$ limit.
However, the process of compactifying the extra spacetime dimension cannot  
change the leading-order divergence structure of  an amplitude;
an amplitude which is finite as $R\to \infty$ must be finite for all values of $R$.
This radius-independence of the leading divergence structure follows from the fact
that the UV behavior of an amplitude should be independent of the large-scale
geometry of our smooth spacetime manifold.
(Indeed, one of the primary alternative regularization methods to be discussed in Sect.~4
will depend on this fact.)

The only missing step, then, is to demonstrate that our five-dimensional
box amplitude
is finite in the $R\to\infty$ limit if the Ward-Takahashi identities hold. 
However, this result is well-known in the four-dimensional case (see, \eg,
Ref.~\cite{Peskin}), and every step of the proof carries directly over
to the case of the one-loop box amplitude in five dimensions.  
The only difference is that rather than having
a degree of divergence of $-4$ (as in four dimensions), this amplitude now has
a degree of divergence of $-3$.
  
One might worry that this proof has a potential loophole.
Since the individual diagrams contributing to the box amplitude
are separately superficially divergent, a bad choice of regulator
could disturb the cancellation between diagrams triggered by gauge invariance,
thereby yielding an incorrect, divergent result.
However, it is always possible to use a gauge-invariant regulator such as 
the Pauli-Villars regulator in order to render each diagram individually
superficially convergent.  There is then no danger of destroying
the cancellations between diagrams, and the Pauli-Villars regulator
can be lifted at the end of the calculation. 
Indeed, this ``pre-treating'' of each diagram with a Pauli-Villars 
regulator can also be used to justify the Furry-theorem cancellations inherent
in the $N=1,3,5$ diagrams.  

Within box diagrams, such cancellations are actually rather robust.
For example, in the four-dimensional case, the required cancellations
are known to occur 
in a special case (so-called ``Delbr\"uck scattering''~\cite{boxpapers}) 
even when a simple hard cutoff is used.  

We thus conclude that the EDR procedure preserves the 
Ward-Takahashi identities for all possible one-loop 
diagrams in five-dimensional QED compactified on a 
circle.

\subsubsection{The fate of $\delta$}

Thus far, we have shown that our momentum integrations and KK sums
must have cutoff parameters $\epsilon$ and $\Lambda$ which are
related through Eq.~(\ref{eps_lam}).  This expression is sufficient
to describe the manner in which $\epsilon$ and $\Lambda$ are correlated
as $\epsilon\to 0$ (or as $\Lambda\to \infty$).  

However, each side
of this relation contains additional terms [${\cal O}(\epsilon)$ and $\delta$
respectively] which vanish in these limits.
Even though these terms individually vanish,
it may seem that determining these terms can be critical 
for performing radiative calculations.  
For example, in a given calculation, $\delta$ may eventually be multiplied 
by terms which grow as $\Lambda\to \infty$;  this structure is already apparent
in expressions such as Eq.~(\ref{therdem}).
Thus, it may appear that $\delta$ can give
rise to non-zero terms which contribute to the final results
of radiative calculations, even after the cutoff is removed.

Clearly, the precise form of the ${\cal O}(\epsilon)$ terms
will depend on the specific diagram in question, much as we expect in
ordinary 4D dimensional regularization.  Consequently, we expect
that $\delta$ will also be a diagram-dependent quantity.
We stress, however, that the {\it relation}\/ (\ref{eps_lam}) 
is itself general.  Indeed, the only diagram-dependence is in 
how certain terms (which vanish as the cutoffs are removed)
are reallocated between ${\cal O}(\epsilon)$
and $\delta$ in Eq.~(\ref{eps_lam}).  

We shall now discuss the fate of $\delta$ as 
a contributing factor in any field-theory calculation.
As we shall explain, no physical observable can possibly
depend on $\delta$.  Therefore, it is never necessary to calculate $\delta$
for any given diagram, and the universal relation in Eq.~(\ref{eps_lam}) 
is sufficient for the calculation of any physical observable.

This claim ultimately rests on the observation that any 
physical observable
must be finite and regulator-independent.
For example, a diagram such as that in Fig.~\ref{fig1}
represents a one-loop mass shift for the external particle.
If $L_n$ represents the value of this diagram 
when the external particle carries KK mode number $n$,
we know that each $L_n$ might individually be divergent;
it is only after {\it renormalization}\/ 
that such a one-loop corrected mass becomes finite.
However, {\it differences}\/ such as $L_n-L_0$ represent one-loop
radiative contributions to the mass differences between different 
KK modes.  Since such mass differences are physical observables,
quantities such as $L_n-L_0$ should be both finite and regulator-independent. 
In an upcoming paper~\cite{paper2},
we shall demonstrate that such differences 
are indeed regulator-independent:  
even though the raw expressions for the loop-diagram differences
appear to contain the regulator cutoffs, these cutoffs can all
be eliminated through resummations and cancellations.
However, imposing the requirement of finiteness on these differences
will lead us to our observation about the irrelevance of $\delta$.

We begin by considering the result of any single diagram.
Our interest is in the behavior of such a diagram as our cutoff is removed
(\ie, as $\Lambda\to\infty$), so we shall concentrate on only those contributions
which potentially survive as $\Lambda\to \infty$. 
In general, following steps such as those which led to
Eq.~(\ref{therdem}), we may express the value of any particular
diagram $L^{(i)}$ in the form
\beq
        L^{(i)} ~\sim~ 
        \alpha_0^{(i)} +  \alpha^{(i)}(\Lambda)  + 
             \delta^{(i)}(\Lambda) \, \beta^{(i)}(\Lambda)~
\label{genform}
\eeq
where the symbol `$\sim$' indicates that we are only
retaining terms which survive as $\Lambda\to \infty$. 
In Eq.~(\ref{genform}), $\alpha_0^{(i)}$ is a diagram-dependent
constant term, while  
$\alpha^{(i)}$ and $\beta^{(i)}$ are diagram-dependent diverging functions of $\Lambda$.
Likewise, $\delta^{(i)}$ is our diagram-dependent $\delta$-parameter.
Even though $\delta^{(i)}$ is assumed to vanish as $\Lambda\to \infty$,
it multiplies a potentially divergent function $\beta^{(i)}(\Lambda)$
and thus can still give rise to a contribution which survives as $\Lambda\to \infty$.
In general, this contribution will take the form
\beq
      \delta^{(i)}(\Lambda) \, \beta^{(i)}(\Lambda)~\sim~
         b_0^{(i)} + b^{(i)}(\Lambda)
\eeq
where once again $b_0^{(i)}$ is a potential constant ($\Lambda$-independent) term
and $b^{(i)}(\Lambda)$ is a divergent function of $\Lambda$.

Given these individual diagrams $L^{(i)}$, the correction to a physical 
observable at one-loop order will always take the form of a linear combination $\sum c_i L^{(i)}$.
Such a physical observable will therefore have the divergence behavior
\beq
    \sum c_i L^{(i)}~\sim~ 
         \sum_i c_i \alpha_0^{(i)} + \sum_i c_i \alpha^{(i)}(\Lambda) +
            \sum_i c_i b_0^{(i)} + \sum_i c_i b^{(i)}(\Lambda)~.
\eeq
However, because this corresponds to a physical observable, we know that this
expression must be finite as $\Lambda\to \infty$.
We therefore have that
\beq
         \sum_i c_i \alpha^{(i)}(\Lambda)  ~=~ - \sum_i c_i b^{(i)}(\Lambda)~.
\eeq
Moreover, as we shall explain below, we further claim that
\beq
          \sum_i c_i b_0^{(i)} ~=~ 0~.
\label{noconstant}
\eeq
Thus, regardless of the precise value of  
the $\delta^{(i)}(\Lambda)$ functions,
we see that their entire purpose is simply to soak up all other potential 
divergences from physically observable quantities.
In the end, the final result for any physical observable in the $\Lambda\to \infty$
limit is given by  $\sum_i c_i \alpha_0^{(i)}$,
and this quantity is completely $\delta^{(i)}$-independent.

Of course, a critical step here was the assumption in Eq.~(\ref{noconstant})
that $\sum_i c_i b_0^{(i)}=0$.
However, this quantity must cancel because  
it is regulator-dependent (depending ultimately on the individual $\delta^{(i)}$'s).
Indeed, as we have discussed above, this quantity is related to the
regulator-dependent ${\cal O}(\epsilon)$ terms through Eq.~(\ref{eps_lam}),
and as such these $b_0^{(i)}$ terms  
are analogous to 
the factors of $\log(4\pi)$ or the Euler-Mascheroni constant $\gamma$
which appear in dimensional regularization calculations but have no
observable effects. 
The cancellation in Eq.~(\ref{noconstant}) is merely the expression
of the fact that such terms will always cancel in the calculation of
any physical observable.

Thus, we conclude that the $\delta$ terms in Eq.~(\ref{eps_lam}) ---
although potentially important for the value of any individual diagram
$L^{(i)}$  --- will ultimately be irrelevant for the calculation of
any physical observable.
Therefore, as indicated above, it is never necessary to calculate $\delta$
for any given diagram, and the universal relation in Eq.~(\ref{eps_lam}) 
is sufficient for the calculation of any physical observable.

\section{Comparisons with Other Regulators 
\label{need}}

In this section, we shall compare our techniques with other regulators
that exist in the literature for dealing with higher-dimensional
quantum field theories with compactified extra dimensions.
We shall pay particular attention to existing
methods which respect higher-dimensional symmetries, with the purpose 
of demonstrating that our regulator successfully reproduces 
results that can be obtained by these methods. 
However, we also shall explain why our particular regulators are useful,
despite the existence of alternatives. 
We shall also illustrate the unwanted complications that can emerge
when one employs a regulator which does not respect higher-dimensional
symmetries.

\subsection{Review of existing techniques \label{exist}}

We begin by reviewing various regularization techniques which have already
appeared in the literature.

The most straightforward way to analyze radiative corrections on extra
dimensions is to decompose our higher-dimensional fields in terms 
of KK modes, and to treat these modes as heavy 4D particles. 
One defines the theory up to
some large but finite cutoff $\Lambda$, and
the Euclidean four-momenta of particles
and their KK masses are assumed to lie below this cutoff, \ie,
\beq
         p_{E}^2 ~\leq~ \Lambda^2~,
\label{4D_cut}
\eeq
and
\beq
           m_{n}^2 ~\leq~ \Lambda^2~,
\label{KK_cut}
\eeq
where $m_n$ is the mass of the $n^{\rm th}$ KK mode.  For compactifications
on a circle, these  masses given by the usual dispersion relation:
\beq
          m_{n}^2 ~=~ m^2 + \frac{n^2}{R^2}~.
\label{kaluza}
\eeq
In the usual treatments, Eqs.~(\ref{4D_cut}) and~(\ref{KK_cut}) are taken 
to be {\it independent}\/ constraints, since such a regulator is insensitive
to the original higher-dimensional nature of the KK theory.
By contrast, the dispersion relation in Eq.~(\ref{kaluza}) is nothing
but the expression of 5D Lorentz invariance which exists at tree level. 

This sort of regulator has been applied in a number of calculations
going all the way back to the original work in Ref.~\cite{DDG}, in which
it was shown that gauge coupling unification can occur with a significantly
reduced GUT scale in a higher-dimensional context, and that large fermion
mass hierarchies can also be generated.   
Since then, regulators such as these have been applied in a 
variety of contexts having to do with precision studies of 
extra dimensions and their diverse effects on ordinary four-dimensional
(zero-mode) physics.

These studies all have one feature in common:
they are concerned with the properties of the zero modes 
and the radiative corrections to 
these properties which are induced by the existence of the excited KK states.
Because the properties of the zero modes are sensitive to only 
four-dimensional symmetries,
regulators which break five-dimensional symmetries
but preserve four-dimensional symmetries
are sufficient for such calculations. 
For example, it is straightforward to demonstrate
that for calculations involving only zero modes,
the sort of 4D regulator defined in Eqs.~(\ref{4D_cut}) and (\ref{KK_cut}) 
and the 5D regulator we introduced in Sect.~\ref{hard}
will yield results whose divergences differ by at most
an overall multiplicative constant.
However, such a constant can be absorbed into the definition of
the cutoff itself (which is particularly ambiguous in a non-renormalizable
theory),
and these effects necessarily vanish as the regulator is removed. 
Thus, both types of regulators will produce identical 
results for all zero-mode calculations.

Unfortunately, such four-dimensional regulators are insufficient
for calculations of the properties of the excited KK 
modes themselves.
Such regulators are therefore also insufficient for calculations
that aim to compare the properties of the excited KK modes (such
as their masses or couplings) with those of the zero modes,
as might be extracted in a collider experiment.
Indeed, as we shall show explicitly in Sect.~\ref{pitfall}, 
such four-dimensional regulators lead to
unphysical artifacts which are difficult 
to disentangle from true, physical effects.

To date, there are very few calculational methods in
the literature which preserve the original higher-dimensional symmetries
that existed prior to compactification.
However, there are three notable exceptions
which we shall now discuss.

First, it can sometimes happen that no regulator is needed,
even in higher dimensions.
For example,
in Ref.~\cite{app_dob}, a practical example of a
regulator-independent calculation in higher dimensions was given. 
Specifically, the authors of Ref.~\cite{app_dob} calculated $g - 2$ for the
muon in a higher-dimensional Standard Model compactified on universal 
extra dimensions.  For the case of a single extra dimension, 
they found that $g - 2$ received only finite
corrections from KK modes at one-loop order.  Of course, no regulator
was needed in this case.  However, they found that such corrections
diverged logarithmically in six dimensions.

Second, it can sometimes happen that a four-dimensional
regulator might itself be sufficient in higher dimensions.
An example of this phenomenon appears in Ref.~\cite{ghil}.
Applying ordinary 4D dimensional regularization, the author of Ref.~\cite{ghil}
showed that it was possible to obtain regulator-independent results for 
QED on a universal extra dimension.  {\it A priori}\/, one would have expected 
an infinite number of counterterms for this theory, due to its
non-renormalizability. 
However, it was shown in Ref.~\cite{ghil} that only a 
counterterm for the electric charge was needed for
describing corrections to the zero-mode coupling at one-loop
order.  Specifically, the author of Ref.~\cite{ghil} calculated the vacuum polarization diagram
$L^{\mu \nu}(p) = \Pi(p^2)(p^\mu p^\nu - g^{\mu \nu}p^2)$ for a photon
zero mode with four-momentum $p$, and found
that the regulator $\epsilon$ canceled in the difference 
$\Pi(p^2) - \Pi(0)$.  Any divergence in a correction to a higher-order 
coupling operator (\eg, the electron magnetic moment) is therefore solely a 
consequence of the charge renormalization. 
Quantities such as $g - 2$ receive finite 
(hence, regulator-independent) corrections.  However,
the author of Ref.~\cite{ghil} showed that this sort of cancellation occurs only at one-loop
order in 5D, and explicitly demonstrated that additional counterterms
are needed when there are two extra dimensions.
Moreover, there was no discussion of vertex corrections, which are
needed for calculating corrections to higher-order operators. 

To the best of our knowledge, there is only one
other regulator which has appeared in the literature which is intrinsically
higher-dimensional and which preserves higher-dimensional Lorentz and gauge symmetries.
This is the regularization method developed in Ref.~\cite{CMS}. 
This method rests upon the observation that
the effects of compactification
should evaporate in the UV limit, and consequently
the UV divergence of a given diagram  
evaluated on a four-dimensional space with a single 
compactified extra dimension should be the same as the
UV divergence of the same diagram evaluated on a five-dimensional
flat (uncompactified) space. 
One can thus extract a finite result from any given loop diagram
in the compactified theory by subtracting the
value of the corresponding diagram in a theory where all 
of the dimensions are infinite.  
In this way, one therefore obtains~\cite{CMS}
a recipe for extracting finite values from loop diagrams which respects
the full higher-dimensional Lorentz invariance as well as 
whatever higher-dimensional gauge invariance might exist.

Operationally,
the technique in Ref.~\cite{CMS} employs a Poisson resummation in order
to recast the sum over Kaluza-Klein momentum mode numbers $n$ 
within a loop diagram on a compactified extra dimension as 
a convergent sum over a ``dual'' set of {\it winding} numbers $w$. 
It turns out that the 
$w=0$ contribution is nothing but the contribution from the corresponding diagram 
evaluated on the uncompactified space.  This ``regularization'' procedure therefore
amounts to transforming to the dual winding-number basis and then
disregarding the contribution from the $w=0$ winding mode.

As an example, 
using this method, the authors of Ref.~\cite{CMS} examined five-dimensional
QED with massless fermions, compactified on a circle.
Although 
the zero-mode photon does not gain a mass as a result of 
four-dimensional gauge invariance, 
it was found that the masses of the excited KK photon modes are 
each shifted by a uniform amount
\beq
         \Delta m^{2}_{n} ~=~ 
      - {e^2\over 2\pi R^2}\, \sum_{w\not=0} \, {2\over |2\pi w|^3}
         ~=~ - \frac{e^2 \zeta(3)}{4\pi^4 R^2}~,
\label{CMS_eq2}
\eeq
where $e$ is the unit of electric charge and where the 
$\zeta$-function represents the winding-number sum:
\beq
         \zeta(n) ~\equiv~ \sum_{w=1}^\infty \, {1\over w^n}~.
\label{zeta_wind}
\eeq
Indeed, most of the results obtained using this method involve the $\zeta$-function
as a sum over winding numbers.

We note that it was strictly for gauge fields that the authors 
of Ref.~\cite{CMS} found such a splitting pattern.
In an upcoming paper~\cite{paper2}, we shall show that such
splittings also occur for other types of particles, even when there is no
gauge symmetry. However, we find that 
these types of splittings occur only when the four-dimensional masses 
of our particles are non-zero (a case which was not considered 
in Ref.~\cite{CMS}).

It is important to note that
the procedure introduced in Ref.~\cite{CMS} 
is {\it not}\/, strictly speaking, a regulator.  
Indeed, a regulator is a way of
temporarily deforming a divergent expression to render it finite;
such deformed expressions are then parametrized by a 
continuous deformation parameter 
(such as $\Lambda$ or $\epsilon$) which is removed at
the end of the calculation.  
For example, let us assume that two expressions $A$ and $B$ are each separately
divergent, but their difference is a physical quantity and therefore finite.
Rather than separately evaluate $A$ and $B$, we might instead evaluate
$A'$ and $B'$, where $A'$ and $B'$ are regulated, finite expressions.
We would then find that $A'-B'$ is either regulator-independent, or tends to
a finite value as the regulator is removed.

By contrast, 
the procedure introduced in Ref.~\cite{CMS} is simply a method 
of extracting a finite expression from a single, infinite diagram. 
In general, we have no assurance that this finite expression corresponds to 
any physical quantity unless the particular calculation we are doing
happens to lead to this expectation for other reasons.
For example, let $L_{n}|_R$ denote the value of a one-loop
vacuum-polarization diagram 
with an external KK photon with mode number $n$,
evaluated when our extra spacetime dimension has radius $R$,
and let 
$L_n|_\infty$ denote the value of the corresponding vacuum-polarization
diagram on an infinite extra dimension. 
(The subscript $n$ in the uncompactified case indicates
that the fifth component of our external photon momentum is still given by $n/R$, 
just as in the compactified case.)
Let us also define $\tilde L_n|_R$ as that portion of $L_n|_R$ which
renormalizes the mass (\ie, $\tilde L_n^{\mu\nu}|_R$ would represent the piece
within $L^{\mu\nu}_n|_R$ which is proportional to the metric $g^{\mu\nu}$).
Within such a setup, we can then write expressions such as $\tilde L_n|_R-\tilde L_0|_R$ in the form
\beq
   \tilde L_n|_R - \tilde L_0|_R ~=~ 
      (\tilde L_n|_R - \tilde L_n|_\infty) - 
      (\tilde L_0|_R - \tilde L_0|_\infty) ~
\eeq
where we have taken $\tilde L_n|_\infty=\tilde L_0|_\infty$ (as occurs when appropriate 
renormalization conditions are applied, such as placing the external photons on-shell
in each case).
Now, the residual four-dimensional gauge symmetry 
requires that $\tilde L_0|_R$ should vanish for all $R$ (including $R\to\infty$),
whereupon we conclude that the physical difference $\tilde L_n|_R-\tilde L_0|_R$ is actually finite and 
given by $\tilde L_n|_R - \tilde L_n|_\infty$. 
Indeed, it is for this reason that this technique is capable of evaluating 
radiative shifts to individual KK masses, even though it was designed
only to yield differences between corrections to quantities in a
compactified theory and an uncompactified one.\footnote{
        Note that in this specific example of KK-photon mass renormalization,
        the above results also imply that $\tilde L_n|_\infty=0$ for all $n$.
        Of course, this can be easily understood as a result of
        {\it five}\/-dimensional gauge invariance.
        Thus, in this particular case, our original diagram $\tilde L_n|_R$
        was already finite by itself, and indeed the subtracted term $\tilde L_n|_\infty$
        vanishes.  We have nevertheless chosen to present this somewhat ``null'' 
        example because this is the original example given in Sect.~II of Ref.~\cite{CMS}.
        In this context, we remark that although the result~\cite{CMS} quoted 
        in our Eq.~(\ref{CMS_eq2}) is correct, it would be incorrect to make the further
        assumption
        that the $w=0$ contribution follows the same functional form as the $w\not=0$ 
        contributions, diverging as $1/w$ with $w\to 0$.  Indeed, as we have 
        explained above, the $w=0$ contribution actually vanishes by five-dimensional 
        gauge invariance, and a direct calculation of the $w=0$ contribution
        will yield an expression which is either identically zero,
        or occasionally indeterminate in the absence of a suitable regulator.}

Even though the method of Ref.~\cite{CMS} is not, strictly speaking,
a regulator, it is nevertheless possible to generalize this method
slightly in order to make it a full-fledged regulator.
For example, we could always write any (divergent) expression $L_n|_R$
in the form
\beq
     L_n|_R ~=~ (L_n|_R - L_n|_\infty) + L_n|_\infty~.  
\label{mixedregs}
\eeq
The first term would then clearly be finite, and the second 
term could be regularized using any of the standard higher-dimensional
regulators 
that apply in an uncompactified space.
Together, we would then have a bona-fide regulator prescription
which could be universally applied for any expression $L_n|_R$.  
However, such a regulator would involve two separate methods,
one for each of the terms in Eq.~(\ref{mixedregs}), and would thus be
relatively awkward to employ in practical settings.

\subsection{Comparisons with previous results
\label{CMS_check}}

If our EHC and EDR regulators are valid, they must reproduce the results
derived via the winding-number technique discussed above.
In this section, we shall show that this is indeed the case.

We first consider the squared-mass shift described by Eq.~(\ref{CMS_eq2}). 
This shift is derived from the part of the vacuum polarization diagram in 
Eqs.~(\ref{kdform1}) and (\ref{kdform2}) which
is proportional to $g^{\mu \nu}$. 
As above, we define $\tilde{L}^{\mu \nu}$ to be this part of the diagram. 
Let us first evaluate this expression following our
extended dimensional-regularization (EDR) procedure.  Utilizing 
our $\Lambda(\epsilon)$ relation in Eq.~(\ref{eps_lam}) 
and explicitly performing the sum over KK modes, we obtain
\beqn
   \tilde{L}^{\mu \nu} & = & 
          -\frac{ie^2 g^{\mu \nu}}{4\pi^2 R^2}\, \lim_{\Lambda R \to \infty} \,
          \Bigg\{ \frac{4}{9}(\Lambda R)^3 - \frac{\Lambda R}{3} - 
          \left[ \frac{4}{3}(\Lambda R)^3 + 2 (\Lambda R)^2 + \frac{2 \Lambda R}{3} \right] \log(\Lambda R) \nonumber \\
    && ~~~~~~~~~~~~~~~~~~~~~~~~+ 2\,\sum_{r = 1}^{\Lambda R}r^2 \, \log(r^2) \Bigg\}~.
\label{mu_nu_eval}
\eeqn
Therefore, our regulator will not reproduce the result in
Eq.~(\ref{CMS_eq2}) unless 
\beqn
         && \lim_{\Lambda R \to \infty} \Bigg\{ \frac{4}{9}(\Lambda R)^3 -
         \frac{\Lambda R}{3} - \left[ \frac{4}{3}(\Lambda R)^3 + 2 (\Lambda
         R)^2 + \frac{2 \Lambda R}{3} \right] \log(\Lambda R) 
                   \nonumber \\ 
          && ~~~~~~~~~~~~ + 2\, \sum_{r = 1}^{\Lambda R}\, r^2 \log(r^2) \Bigg\}
          ~~~  {\stackrel{\displaystyle ?}{=}}~~~  \frac{\zeta(3)}{\pi^2}~.
\label{sum_id}
\eeqn
On the surface, such an identity would seem somewhat improbable, since
the left side involves individual terms which are each manifestly divergent,
while the right side is finite.
Indeed, some of the terms on the left side of Eq.~(\ref{sum_id}) 
are simple polynomials in $\Lambda R$,
while the expression on the second line is a discrete sum in which $\Lambda R$ appears
as an upper limit. 

Surprisingly, however, it is easy to verify numerically that Eq.~(\ref{sum_id})
holds to any precision desired.
Indeed, the expression on the left side of this identity experiences
a remarkably fast convergence to  
$\zeta(3)/\pi^2$,  already coming within
$10\%$ of this value for $\Lambda R=1$, and coming within $1\%$ 
for $\Lambda R=9$.
In fact, Eq.~(\ref{sum_id}) is an entirely novel mathematical representation for
the $\zeta$-function as the
limit of an infinite summation.
Equivalently, inverting this relation provides an analytical form for the infinite
sum $\sum_r r^2 \log(r^2)$, which can be useful in many contexts
dealing with KK summations.

This, then, provides a highly non-trivial check of our extended
dimensional regularization (EDR) procedure.
By demonstrating that EDR is consistent with
the technique in Ref.~\cite{CMS}, we once again verify that 
EDR indeed preserves both
higher-dimensional Lorentz invariance and higher-dimensional
gauge invariance, as promised. 
Although we have only shown a comparison for one particular diagram,
it is straightforward to verify that similar cross-checks hold
for  other diagrams as well.

We can also verify that our extended hard cutoff (EHC) regulator
is consistent with the method of Ref.~\cite{CMS}.
However, in order to make such a comparison, we should examine
a theory which exhibits higher-dimensional Lorentz invariance
but not higher-dimensional gauge invariance.

For this purpose, let us examine a toy five-dimensional model
consisting of a single scalar $\phi$ and a single fermion $\psi$ compactified 
on a circle and experiencing a Yukawa interaction of the form
$G \phi (\overline{\psi}\psi)$ where $G$ is the five-dimensional Yukawa coupling.
Indeed, this theory will be analyzed more extensively in Ref.~\cite{paper2}.
Within this theory, let us examine the  
one-loop diagram which renormalizes the squared mass of a 
KK excitation of the scalar field with mode number $n$.
This diagram is shown in Fig.~\ref{fig1}, where we now take
the external lines to represent KK modes of the scalar $\phi$
and the internal lines to represent KK modes of the fermion $\psi$.
As before, we shall write $L_{n}|_R$ to denote
the value of this diagram 
when our extra spacetime dimension has radius $R$, and we
shall write $L_n|_\infty$ to denote the corresponding diagram
on an infinite extra dimension. 
Note that in the latter case, despite the disappearance of discrete
KK modes, the subscript $n$ continues to be specified 
as a reminder that the fifth component of the external momentum 
in such a diagram should continue to carry the value $n/R$.

Because gauge invariance is not a symmetry of this theory, 
it will be sufficient
to employ our extended hard cutoff (EHC) regulator in evaluating this diagram.
Following the procedure outlined in Sect.~\ref{hard}, 
we then obtain the expression
\beqn
   L_n|_R - L_n|_\infty & = & \frac{ig^2}{4\pi^2 R^2}
        \, \lim_{\Lambda R \to \infty} \, \Bigg\{ \frac{4}{9}(\Lambda R)^3 -
         \frac{\Lambda R}{3} \nonumber \\ 
    && ~~~- \left[ \frac{4}{3}(\Lambda R)^3
         + 2 (\Lambda R)^2 + \frac{2 \Lambda R}{3} \right] \log(\Lambda R) +
   2\, \sum_{r = 1}^{\Lambda R}\, r^2 \log(r^2) \Bigg\}~. \nonumber \\
\label{our_yuk}
\eeqn
The quantity $\Lambda$ now represents our hard cutoff, which is 
the same for both diagrams, and $g\equiv G/\sqrt{2\pi R}$ represents the Yukawa
coupling of each individual KK mode.  
Note that the above result holds for any value of $n$, including $n = 0$,
and holds  independently of whether the 5D scalar is real or 
complex (since the scalar does not run in the loop). 
By contrast, the regularization technique of 
Ref.~\cite{CMS} leads to the expression
\beq
        L_n|_R - L_n|_\infty ~=~ \frac{ig^2}{4\pi^4 R^2}\, \zeta(3) 
             ~=~  \frac{iG^2}{8\pi^5 R^3}\, \zeta (3)~.
\label{CMS_yuk}
\eeq
However, once again, the identity
in Eq.~(\ref{sum_id}) ensures that these results are equivalent.
Indeed, we see that Eq.~(\ref{sum_id}) essentially serves as a mapping
between the results derived using the methods of this paper and
those derived using the methods of Ref.~\cite{CMS}.

Although these UV regulators yield the same results
for mass corrections, they
nevertheless treat {\it infrared}\/ (IR) divergences 
differently. 
Because there is no direct relationship between the IR
divergence that results in a given diagram when an extra dimension is
compactified and the IR divergence that results when the extra dimension
is infinite, the regularization method of Ref.~\cite{CMS}
does not eliminate IR divergences.
Indeed, the discrete KK sum that results for a compactified extra dimension
and the KK integral that 
would result in the case of an infinite dimension only become
more dissimilar in the IR limit. 
Of course, the regulators in this paper also leave IR divergences intact.
However, because the method of Ref.~\cite{CMS}
requires that we pass from a KK momentum basis
to a KK winding basis in order to eliminate the UV divergence,
any IR divergence which remains is redistributed across all winding modes,
particularly those with large winding numbers, and
can no longer easily be isolated.  By contrast, because our methods
do not require any such reorganization, the IR divergences that
remain in our method continue to be easily identified and treated.

As a concrete example of these ideas,
let us consider the vacuum polarization diagram $L^{\mu \nu}_n$
in the case in which the external KK photon of mode number $n$ is on-shell 
and the bare (five-dimensional) mass $M$ of the fermion running in the loop is zero.
Using our EDR procedure, we obtain the results
in Eqs.~(\ref{kdform1}) and (\ref{kdform2}).
Although the integrands in Eq.~(\ref{kdform2}) are finite for each non-zero $r$,
the quantity $W$ in Eq.~(\ref{kdform3}) diverges for $n=r=0$, \ie, for a zero-mode
external photon with a zero-mode
fermion running in the loop.
This is the IR divergence, encapsulated entirely within the zero-mode 
contribution to the KK sum in Eq.~(\ref{kdform1}).
By contrast, if we were to use the methods of Ref.~\cite{CMS} to analyze
the same vacuum-polarization diagram, we would obtain the result
\beq
       L^{\mu \nu}_0|_R - L^{\mu \nu}_0|_\infty ~=~
       \frac{ie^2}{4\pi^2}\frac{p^{\mu}p^{\nu}}{3} \, \sum_{w \neq 0}\, \frac{1}{|w|}~.
\label{IR_div}
\eeq
In this case, the IR divergence is reflected in the divergence
of the winding-number sum, and cannot be isolated to a particular term
within Eq.~(\ref{IR_div}).

Note that IR divergences
can generally be regularized through the introduction of small masses.
For example, the IR divergence discussed above
is eliminated when the fermion is given a small non-zero four-dimensional mass 
or the external photon is slightly off-shell. 
The introduction of such a mass is relatively straightforward to 
implement within the framework of the regulators in this paper.
However, the introduction of such a mass within the framework
of Ref.~\cite{CMS} might be significantly more complicated. 
Such an IR regulator would inevitably be redistributed across
every contribution to the winding-number sum (rendering it finite), 
but such a sum is not likely to have a simple mathematical form.
Alternatively, one could imagine regulating a sum such as that
in Eq.~(\ref{IR_div}) directly (e.g., by inserting a small
Boltzmann-like suppression factor), but such an insertion
is likely to break higher-dimensional Lorentz invariance
or gauge invariance.  Moreover, it is not clear that transforming
such a factor back to the KK momentum basis would provide it 
with any clear physical interpretation.

We have seen, then, that the regulators we have proposed in this paper
are able to reproduce the corresponding results of Ref.~\cite{CMS} when
appropriate.
However, to be truly useful, our techniques
also must apply in situations where other methods do not. 
Since the technique in Ref.~\cite{CMS} operates
strictly in the winding-number basis, it loses information about
contributions to radiative corrections from different physical momentum scales.
This poses no problem in calculations of radiative
corrections to physical parameters (\eg, masses and couplings) which
would be observed in experiments. However, it is not possible to
calculate Wilsonian renormalization-group evolutions of such parameters in this
scheme. If extra dimensions are discovered at a future collider, it
may be desirable to define EFT's for KK modes below the center-of-mass
(CM) energy. Calculating the parameters in such a theory would require
the use of the Wilsonian renormalization group, with the corresponding evolution of
parameters running from the UV to the CM energy. As we shall see in Refs.~\cite{paper1b,paper2},
our regulators can handle such calculations explicitly.
Indeed, this was one of our original motivations for developing the new regulators
in this paper.

\subsection{The necessity of preserving higher-dimensional Lorentz invariance 
\label{pitfall}}

In this section, we illustrate the pathologies which appear when using
regulators that break higher-dimensional Lorentz invariance. 
As a concrete example, we shall again consider
our toy five-dimensional model
consisting of a single scalar $\phi$ and a single fermion $\psi$ compactified
on a circle and experiencing a Yukawa interaction of the form
$G \phi (\overline{\psi}\psi)$ where $G$ is our five-dimensional Yukawa coupling.
Within this theory, we shall attempt to calculate the radiative corrections to 
the KK masses of the scalar using a regulator which preserves four-dimensional
Lorentz invariance but breaks five-dimensional Lorentz invariance.

Once again, we shall do this 
by calculating the difference between a loop diagram
which renormalizes the squared mass of a scalar mode in Yukawa theory
and the corresponding diagram for the zero mode.  We define $L_n (p)$
to be the squared-mass renormalization diagram for a scalar with
mode-number $n$ and four-momentum $p$ (shown in Fig.~\ref{fig1}). For
simplicity, we take the zero-mode masses $m_{\psi}$ and $m_{\phi}$ of
these two fields to vanish.  We then find
\beq
  L_n ~=~ 4ig^2 \, \int_{0}^{1}dx\, \sum_r \, \int \frac{d^4\ell_E}{(2\pi)^4}\,
      \left[ \frac{\ell_{E}^{2} + r(r - n)/R^2 + x(1 -x)n^2 /R^2}
                  {(\ell_{E}^{2} + (r - xn)^2 /R^2)^2}
           \right]
\label{scagram}
\eeq
where $g\equiv G/\sqrt{2\pi R}$. 
Note that we now write $(r - xn)/R$ rather than $\ell^4$ 
because we are no longer treating this quantity as 
the fifth component of a five-vector.

The expression in Eq.~(\ref{scagram}) is badly divergent, and must be
regularized.  Let us therefore place a 4D cutoff $\Lambda$ on $\ell_E$ and truncate
the KK sum at this cutoff.  In other words, we shall take our
integration limits to be given by $\ell_{E}^{2} \leq \Lambda^2$
and our summation limits to be given by $-\Lambda R \leq r \leq \Lambda R$.
Note that these constraints break higher-dimensional Lorentz
invariance, since they separately regularize four-momentum integrals
and KK sums. 
Nevertheless, imposing these constraints, we find that
\beqn
   L_n - L_0 & = & -\frac{ig^2}{4\pi^2 R^2}\,\sum_{r = -\Lambda R}^{\Lambda R}\,
              \int_{0}^{1}dx \, \Bigg\{ (-2x^2 + x)n^2\nonumber\\
    && ~~~~~+ ~\frac{ (r - xn)^4 + (r - xn)^2 \left[ x(x - 1)n^2 - r(r - n) \right]}
            {\Lambda^2 R^2 + (r - xn)^2} \nonumber\\
    && ~~~~~+ ~\left[ 2(r - xn)^2 + x(x - 1)n^2 - r(r - n) \right]  \nonumber \\ 
    && ~~~~~~~~~~\cdot ~\Big[ \log( \Lambda^2 R^2 + (r - xn)^2) - 2\log(r - xn) \Big] \nonumber \\ 
    && ~~~~~- ~r^2  \left[ \log( \Lambda^2 R^2 + r^2) - 2\log r \right] \Bigg\}~.
\label{L_cut}
\eeqn
Clearly, this expression diverges linearly with $\Lambda$. 
This is a problem, since this quantity corresponds to the difference 
between squared masses, which should be finite. 

The reason this divergence appears is that
the loop diagrams in this equation do not determine renormalized
masses by themselves.  Rather, each KK mode should have a counterterm for its
squared mass, and  a calculation of a squared mass difference must include
these counterterms.  Such terms would indeed cancel artificial violations of
Lorentz invariance.  However, they also would break the KK dispersion
relation for the underlying theory, since they are part of the bare
Lagrangian.

This situation has an analogue in four-dimensional QED. 
If we use a hard cutoff to regularize divergences in that theory,
we then generate a photon mass which is proportional to the cutoff. 
As well as being divergent, such a mass term violates gauge
symmetry.  However, as is well known (see, \eg, Ref.~\cite{Peskin}),
this problem can be remedied by introducing
counterterms which break gauge invariance and cancel the unphysical
effects from loop diagrams. 
However, our bare Lagrangian is then no longer gauge invariant.

In 5D Yukawa theory, the relevant symmetry is higher-dimensional
Lorentz invariance.  In the spirit of QED, it may therefore appear
straightforward to introduce counterterms to cancel regulator-induced
violations of 5D Lorentz invariance.  However, 
the compactification of an extra dimension also breaks higher-dimensional Lorentz 
invariance at finite scales.  This violation can manifest itself 
in an EFT as a violation of the usual 5D dispersion relation, as in the 
case of Eq.~(\ref{CMS_eq2}). 
Therefore, counterterms would not only have to cancel
unphysical violations induced by our regularization scheme, but
nevertheless preserve the bona-fide effects induced by the compactification itself. 
Without {\it a priori} knowledge of what the results should be, it would be
quite difficult to determine which effects would be physical and which would not.
Indeed, it would be difficult to deduce the form of appropriate counterterms 
if we limit ourselves to this sort of regulator. 
Such a regulator, therefore, does not lend itself to a straightforward
calculation involving the relative renormalizations of the parameters describing
a KK spectrum.

As required, the regulators developed in this paper yield finite
loop-diagram differences and thus avoid this problem. 
We therefore did not need to introduce counterterms, since the squared masses of
KK states --- which are renormalized by our loop diagrams --- all carry
the same divergence at tree level. Indeed, the dimensionless squared masses
are given by the relation $m_{n}^{2}R^2 = m_{0}^{2}R^2 + n^2$ at tree level, 
and only the $m_{0}^{2}R^2$-term diverges in the UV.~ 
Hence, only one counterterm is needed for the entire mass spectrum 
of KK states, and the effects of such a counterterm  
cancel when calculating {\it differences} between squared masses. 
Similar results hold for other types of loop diagrams. 
It is for this reason that our 
techniques can produce regulator-independent EFT's.
These issues will be discussed in more detail in Ref.~\cite{paper1b}.

\section{Conclusions and Future Directions
\label{conclude}
}
\setcounter{footnote}{0}

In this paper, we proposed two new regulators (EHC and EDR) for quantum
field theories in spacetimes with compactified extra dimensions.  
Although they are based on traditional four-dimensional regulators,
the key new feature of these higher-dimensional regulators is that 
they are specifically designed to handle mixed spacetimes in which some
dimensions are infinitely large  and others are compactified. 
Moreover, unlike most other regulators which have been used in 
the extra-dimension literature, these regulators are designed to 
respect the original higher-dimensional Lorentz and gauge symmetries 
that exist prior to compactification, and not merely the four-dimensional 
symmetries which remain afterward.  

As we have discussed, 
these regulators should be particularly useful for
calculations of the physics of the excited Kaluza-Klein modes in any
higher-dimensional theory, and not merely the radiative effects that these excited
KK modes induce on zero modes.  Indeed, by
respecting the full higher-dimensional symmetries, our regulators
avoid the introduction of spurious terms which would not have been
easy to disentangle from the physical effects of compactification. 

Moreover, as part of our work, we also derived a number of 
ancillary results.  For example, in gauge-invariant theories, we demonstrated
that analogues of the Ward-Takahashi identity hold not only for 
the usual zero-mode (four-dimensional) photons, but for all 
excited Kaluza-Klein photons as well.  

Clearly, the analysis we have done in this paper only begins to
scratch the surface of what is possible.
For example, this analysis has been restricted
to five dimensions and, in many places, to one-loop amplitudes.
While this clearly covers the most pressing situation that
might emerge if extra dimensions are ultimately discovered, it
would be interesting to extend our discussion to multi-loop
amplitudes (where appropriate) and to even higher dimensions.
In particular, 
both of these extensions would involve additional KK sums which
would require their own cutoffs, and thus there will be additional
balancing constraints that must be imposed between these
cutoffs and the regulator of the four-dimensional momentum integral
in order to preserve higher-dimensional Lorentz and gauge symmetries.

Other sorts of extensions are also possible.
For example, in more than five dimensions, we can consider
compactifications not just on flat spaces (such as we have considered
here), but also spaces with their own intrinsic curvatures or warpings.
Moreover, even for flat compactification manifolds, there remains
the possibility of having non-trivial {\it shape}\/ moduli~\cite{shape}.
All of these possibilities represent different types of mixed spacetimes
which would have unusual KK spectra and which would in principle require 
their own analysis. 

There are also other important geometric extensions to consider, even in five dimensions.
For example, although the analysis of this paper has been restricted
to compactification on a smooth manifold, it is important to
extend these results to orbifolded spacetimes which contain boundaries 
(\ie, branes, or orbifold fixed points).
Indeed, compactification on such orbifolded geometries is ultimately required in 
order to obtain a chiral theory in four dimensions.  
In such theories, some processes are purely four-dimensional (occurring
on the branes) while others are five-dimensional and others are mixed.
Although the existence of brane-kinetic terms~\cite{kinetic} can have a profound
effect on the physics on the brane, we nevertheless expect our higher-dimensional
Ward identities to be preserved in the bulk.  Regulators such as those we have developed
here should therefore continue to have application for the bulk physics
in such situations.  
This will be discussed in more detail in Ref.~\cite{paper2}.

Even within the framework of compactification of a single
extra dimension on a circle, there remain important extensions  
of our work which we have not considered. 
For example, we have primarily focused on abelian gauge theories
and their associated Ward identities, but we have not considered
their non-abelian extensions.  
This will be important for ultimately
calculating radiative corrections within, say, a higher-dimensional Standard Model. 
Likewise, in this paper we often considered five-dimensional 
QED.~Although this theory is non-renormalizable, we restricted our attention
to the usual electron/photon coupling and did not allow 
allow additional non-renormalizable interactions.
Even though such interactions should continue to respect our
higher-dimensional Lorentz and gauge symmetries (therefore requiring
the use of a regulator such as we have developed here),
the existence of such interactions can be expected to lead to 
complications beyond those considered in this paper.

Finally, it should be stressed that this work focused on only
one rather narrow type of regulator, namely one in which our KK
sums were regulated through a hard cutoff $\Lambda$.
However, other types of regulators are possible.
For example, an infinite KK sum might alternatively
be regulated through the introduction of Boltzmann-like
suppression factors, \eg,
\beq
      \sum_r \, {1\over r} ~\longrightarrow~ \sum_r \, {1\over r}\, e^{-y |r|}
\eeq
where $y>0$ is a regulator parameter.  One would then take
$y\to 0$ at the end of the calculation, while simultaneously 
maintaining a certain relation between $y$ and $\epsilon$ (analogous
to our EDR relation between $\Lambda$ and $\epsilon$)
so that five-dimensional Lorentz and gauge invariance are maintained.
However, it is not clear what physical interpretation might be
ascribed to such a regulator parameter $y$.
Similarly, we again mention the possibility of preserving 
gauge invariance even with a hard cutoff, but with suitable counterterms 
as well.  However, such counterterms will necessarily break the original
higher-dimensional symmetries of our bare Lagrangian. 

Another approach, first advanced in Ref.~\cite{Nibbelink},
is to rewrite the KK sum as a contour integral in which the different
terms of the sum emerge from the poles of the integrand.
One can then apply a regularization procedure akin to 't~Hooft-Veltmann
dimensional regularization to the integral~\cite{Nibbelink,Contino}.
However, this still results in two independent regulators, one for
the KK integral and another for the four-momentum integral, and five-dimensional
symmetries will generally not be protected unless these two regulators
are balanced in a manner similar to what we have outlined in this paper.

There are, of course, other potential methods of deforming our
KK summations.  For example, we might Poisson-resum our KK summation,
and attempt to apply one of the above regulators to the Poisson-resummed
version instead.  Note that Poisson resummation of the KK sum
was originally introduced into the large extra dimension context 
in Ref.~\cite{Antoniadis}.
There are also other techniques which might be employed, such as proper-time regulators,
zeta-function regularization, {\it etc}\/.  Indeed, these methods ultimately play various roles 
in the different approaches sketched here.
Other approaches towards treating the KK summation
based on dimensional regularization have also been utilized in various calculations~\cite{moreghil}.

Another possibility might be to employ a so-called ``mixed propagator''
formalism~\cite{propagatormethod}.
In such a formalism, the four large dimensions are treated
in momentum space, as usual, while the compactified extra
dimension is treated in position space.  This avoids the
introduction of a KK sum altogether.
However, in such situations the higher-dimensional divergences
are not eliminated --- 
they are the same as would appear in the corresponding
higher-dimensional {\it uncompactified}\/ theory, as this formalism
makes abundantly clear.
This formalism thus lends itself naturally to the treatment
given in Ref.~\cite{CMS}.

Of course, it is possible that the true UV limit of a given higher-dimensional
theory is not higher-dimensional at all~\cite{deconstruction}.
Such ``deconstructed'' extra dimensions would change the UV divergence structure 
of the theory in a profound way that would eliminate the need for many of these
different regularization techniques. 
Indeed, deconstruction can also be used as an alternative  technique 
for performing many of the sorts of radiative calculations for excited KK 
modes that have been our focus in this paper~\cite{pokorski}.
Similarly, radiative corrections may be finite in cases in which
there exist additional symmetries (either unbroken or softly broken)
to protect against divergences.  A well-known example of this would
include radiative corrections in theories with supersymmetry
broken through the Scherk-Schwarz mechanism~\cite{SS} (leading to 
so-called ``KK regularization'', in which the full KK summations lead to
finite results), or in theories
in which the Higgs is identified as a component of a higher-dimensional
gauge field and consequently has a mass for which radiative corrections
are protected by gauge symmetries~\cite{Hosotani}.

Likewise, such higher-dimensional theories may ultimately be embedded
into string theory.  String theory provides entirely new methods of
eliminating divergences which transcend what is possible in quantum
field theories based on point particles. 
Indeed, there even exist several string-inspired methods of 
regularizing field theories directly~\cite{missusy,Ghilencea2,Ghilencea3}.

Another possibility is to retain the full higher-dimensional space but
take a non-perturbative approach towards
extracting exact solutions for the excited KK masses and couplings.
Ideas in this direction have been advanced, \eg, in Ref.~\cite{seibergwitten}.
 
In this connection, it might seem strange that we have not discussed
the Pauli-Villars (PV) regulator.  Indeed, such a regulator
preserves both Lorentz invariance and gauge invariance, even in higher dimensions,
and may be more than sufficient for certain calculations 
(see, \eg, Refs.~\cite{agashe,Contino2}).
However, there are several reasons why such a regulator may not ultimately
be suitable for general calculations in mixed spacetimes, especially
those focusing on the radiative corrections to the properties of the excited
KK modes.
First, the PV regulator does not preserve non-abelian
gauge symmetries, even in four dimensions.
Second, even for the abelian theories which have been our main focus
in this paper, compactification introduces a major algebraic problem:
the PV regulator parameter $\Lambda$ 
becomes inextricably entangled in our KK mode-number sum
except in particular situations (see, \eg, Ref.~\cite{Contino2}) in which
the radiative corrections are already known to be finite. 
Thus, this regulator is particularly unsuited for the mixed spacetimes  
which have been our main focus in this paper.
Of course, it might seem that such a PV  
regulator might nevertheless be suitable for numerical studies which
do not require closed-form analytical expressions.
However, even this is not possible, because
there is a third complication:  {\it unitarity}\/ is not 
preserved using a PV regulator unless the regulator 
parameter $\Lambda$ is sent to infinity.
Thus, it is likely to be difficult to treat such a system 
numerically with any confidence
when our PV regulator is in force.

By contrast, the regulators we have developed in this paper
are designed to be relatively straightforward, intuitive, and easy   
to use for practical calculations.
Indeed, as mentioned at the end of the Introduction,
this paper is only the first in a two-part series. 
In a subsequent companion paper~\cite{paper1b},
we shall discuss how these new regulators may be employed
in order to derive regulator-independent effective field theories at different
energy scales.
We shall also discuss how these regulator techniques can be used
to extract finite results for physical observables that
relate the physics of excited KK modes to the physics of KK zero modes.
Moreover, in a third paper~\cite{paper2},
we shall study the manner in which the 
KK masses and couplings in various higher-dimensional
effective field theories evolve as functions of energy scale,
and as extra spacetime dimensions are slowly integrated out in passing
from the UV to the IR.
In particular, in Ref.~\cite{paper2}, we shall study how the well-known tree-level
relations amongst the tower of KK masses and amongst their couplings
are ``deformed'' when radiative effects are included.
In each case, we shall see that it is the regulators we have
developed here which will enable these calculations to be performed.

\bigskip

\section*{Acknowledgments}
\setcounter{footnote}{0}

We are happy to  thank Z.~Chacko, D. Toussaint, and U. van Kolck
for numerous conversations and insights.
We are particularly grateful to Z.~Chacko for comments on a preliminary
draft of this paper.
This work is supported in part by the National Science Foundation
under Grant PHY/0301998, by the Department of Energy under 
Grant~DE-FG02-04ER-41298, and by a Research Innovation Award from
Research Corporation.

\bigskip

\bibliographystyle{unsrt}

\end{document}